\documentclass[aps,showpacs,superscriptaddress,floatfix]
{revtex4}
\usepackage[dvips]{graphicx}
\newcommand{\beq}{\begin{equation}}
\newcommand{\eeq}{\end{equation}}
\newcommand{\beqa}{\begin{eqnarray}}
\newcommand{\eeqa}{\end{eqnarray}}
\newcommand{\beqan}{\begin{eqnarray*}}
\newcommand{\eeqan}{\end{eqnarray*}}

\newcommand{\affA}{%
    National Institute of Information and Communications Technology,
    4-2-1 Nukuikita, Koganei, Tokyo 184-8795, Japan}
\newcommand{\affB}{%
    CREST, Japan Science and Technology Corporation,
    3-13-3 Shibuya, Tokyo 150-0002, Japan}
\newcommand{\affC}{%
     Tohoku University, 
     Researach Institute of Electrical Communication 
     2-1-1 Katahira, Aoba-ku, Sendai, Miyagi 980-8577, Japan}
\newcommand{\affD}{%
     Department of Physics, 
     Tokyo Institute of Technology 
     Meguro-ku, Tokyo 152-8551, Japan}

\begin{document}
\title{Theory of multiwave mixing 
and decoherence control in qubit array system}
\author{Masahide Sasaki}
\affiliation{\affA}
\affiliation{\affB}
\author{Atsushi Hasegawa}
\affiliation{\affA}
\affiliation{\affB}
\author{Junko Ishi-Hayase}
\affiliation{\affA}
\author{Yasuyoshi Mitsumori}
\affiliation{\affC}
\author{Fujio Minami}
\affiliation{\affD}
\email{e-mail:psasaki@nict.go.jp}
\begin{abstract}
We develop a theory to analyze the decoherence effect 
in a charged qubit array system with 
photon echo signals in the multiwave mixing configuration. 
We present how the decoherence suppression effect 
by the {\it bang-bang} control with the $\pi$ pulses 
can be demonstrated in laboratory 
by using a bulk ensemble of exciton qubits and 
optical pulses whose pulse area is even smaller than $\pi$. 
Analysis is made on the time-integated multiwave mixing signals 
diffracted into certain phase matching directions
from a bulk ensemble. 
Depending on the pulse interval conditions, 
the cross over 
from the decoherence acceleration regime
to the decoherence suppression regime, 
which is a peculiar feature   
of the coherent interaction between a qubit and 
the reservoir bosons, 
may be observed in the time-integated multiwave mixing signals 
in the realistic case 
including inhomogeneous broadening effect. 
Our analysis will successfully be applied 
to precise estimation of the reservoir parameters 
from experimental data 
of the direction resolved signal intensities 
obtained in the multiwave mixing technique.  
\end{abstract}
\pacs{03.67.Pp, 42.50.Md, 78.20.Bh}
%
%
\date{\today}
\maketitle

\section{Introduction}
\label{intro}
The analysis of decoherence of elementary excitations 
in solid state 
is one of the main approaches to understand physics in 
condensed matter. 
It also attracts much attention in device physics 
attempting to implement quantum information processing devices. 
A central issue is how to realize and control coherent evolution 
of well coded quantum states. Effective means should be chosen 
depending on the way of coupling to the system and decoherence 
mechanism. 
Among various proposals 
semiconductor devices have potential advantages, 
including the ease of integration 
and 
the use of matured industrial technologies.

In particular, 
excitons in semiconductor make clear distinction 
from other two level systems for qubits 
in the sense that it can efficiently couple to photons, 
which are the signal carriers to build a communication network. 
In addition, 
its large dipole moment makes possible direct optical control 
of qubits in a time scale of femto second. 
Various results of optical coherent manipulation of exciton 
in semiconductor have been reported, 
such as indirect 
\cite{Cundiff94,Giessen98} 
and direct 
\cite{Schulzgen99} 
observations of excitonic Rabi oscillations 
in higher dimensional semiconductor structures, 
manipulations of a one-qubit rotation 
of a single quantum dot exciton 
\cite{Bonadeo98,Toda00,Stievater01,Kamada01,Htoon02}, 
direct observations of excitonic Rabi oscillations 
in quantum dot ensemble 
\cite{Borri02}, 
and entanglement manipulation in quantum dots 
\cite{Chen00,Bayer01}. 
They open great opportunities for quantum information processing 
based on exciton qubits.

On the other hand, 
exciton qubits always suffer from unwanted coupling with 
the external degrees of freedom of charged excitations 
including the phonon scattering and the scattering by 
excitons themselves, 
leading to the decoherence. 
In particular, 
the fast decoherence time of excitons 
which is typically a few tens of pico second 
will be the first obstacle to be overcome.

The well known methods to fight the decoherence are 
quantum error correcting codes 
\cite{Shor95,Steane96,CalderbankShor96,Gottesman96,KnillLaflamme97,Calderbank_etal97}
and 
encoding with decoherence free subspace 
\cite{PalmaSuominenEkert96,Zanardi97,Duan98,Lidar98,Knill00,Kempe01,Lidar01}
. 
Both methods rely on encoding with multipartite entangled states 
for correcting errors by redundant qubits, 
or 
for making the state evolution immune to decoherence 
by exploiting symmetries of the system-environment interaction. 
It is yet very challenging to prepare the multipartite entangled 
states in the desired form in solid state device.

The other, and only plausible method with present technologies 
might be the dynamical decoupling control by driving the system 
continuously with external pulsed field sequences,
the so called {\it bang-bang} (BB) control 
\cite{Ban98,ViolaLloyd98,Duan99,Viola99a,Zanardi99,Vitali99,Viola99b,Viola00,Vitali02,Byrd01,Cory00,Agarwal01,Uchiyama02}, 
or 
with projecting onto a certain measurement basis 
\cite{Facchi00,FacchiPascazio02,Nakazato03}, 
the so called quantum Zeno effect 
(for recent review, see 
\cite{HomeWhitaker_Rev_Q_Zeno97,FacchiPascazio_Rev_Q_Zeno01}). 
This does not basically require complicated qubit encoding, 
but effective at a single qubit level. 
The former is easier from the technical viewpoint, 
and therefore to be demonstrated as the first step for 
solid state qubits. 
The BB control relies on a time reversal of the decoherece process 
in a short time scale within the reservoir correlation time. 
Similar ideas have been known such as 
spin echoes 
\cite{Hahn50}, 
photon echoes 
\cite{Kurnit64}, 
wave packet echoes 
\cite{Buchkremer00}, 
and charge echoes 
\cite{Nakamura02,Creswick04}. 
These techniques have been powerful tools to investigate 
transient phenomena. 
The BB control can be regarded as a variant of them, 
focusing on an active decoherence control of qubit. 
This has been demonstrated recently 
with a solid-state nulear qubit 
\cite{Ladd04}.

Our concern here is the BB control with optical pulses 
for exciton qubits in semiconductor.  
In order to put this method to the experimental test, 
we will face mainly two problems. 
One is the difficulty of detecting the expected 
stabilization effect by weak signals from a single exciton 
qubit. 
The other is that 
the presently available laser power is not strong enough yet 
for preparing the $\pi$ pulse area. 
One possible solution is to use a bulk ensemble of excitons. 
Strong signals from the ensemble make the detection easy. 
Furthermore they make it possible to observe the decoherence 
suppression effect due to the ideal $\pi$ pulses 
even if the irradiated pulses have the area of smaller than 
$\pi$. 
This is because some fraction of qubits can effectively 
feel the ideal $\pi$ pulses, 
and they form a macroscopic polarization lattice 
which radiates photon echo signals in a certain phase 
matching direction. 
The bulk ensemble is, however, always associated with 
irrelevant phenomena to the present issue, 
such as inhomogeneous broadening of the energy levels 
and complicated light diffraction due to 
polarization lattices of excitons.

It is then necessary to extend the theories for a single 
qubit to a qubit array system to analyze the effect of 
our interest with rather complicated signals. 
The purpose of this paper is to develop a theory 
to analyze the decoherence with the photon echo signals 
in the multiwave mixing configuration, 
and 
to predict the effects of decoherence control 
in a practical experimental setting.

The paper is organized as follows. 
In section 
\ref{Model} 
we present the model and introduce basic notations. 
In section 
\ref{State evolution under pi pulses} 
we first develop a formulation 
to evaluate the multiwave mixing signals 
in the ideal $\pi$ pulse case. 
We then present the results in the case of homogeneous 
broadening, 
and introduce basic notions to analyze 
the state evolution of the qubit-reservoir system 
using the multiwave mixing signals. 
In section 
\ref{State evolution under weak pulses} 
we develop a formulation for the weak pulse case, 
and present numerical simulations on 
the time-resolved 
and time-integated multiwave mixing signals. 
They include not only the suppression but also 
the acceleration of decoherence 
induced by the optical BB control. 
Section  
\ref{Concluding remark} 
is for concluding remark.

\section{Model}
\label{Model}

In this paper we consider an ensemble of qubits 
and identical reservoirs.  
The model Hamiltonian is given by 
\beq  
\hat H(t)=\hat H_0 + \hat H_{\rm QR} + \hat H_{\rm QF}(t), 
\eeq
where
\beq
\hat H_0=\sum_j
        \left[
          {1\over2}\hbar\nu_j \hat\sigma_z(j)
        + \sum_l\hbar\Omega_l \hat B^\dagger_l(j) \hat B_l(j)
        \right], 
\eeq
\beq
\hat H_{\rm QR}
=
\sum_j\hbar\hat\sigma_z(j)
\sum_l
   \left[
     g_l \hat B^\dagger_l(j) + g^\ast_l \hat B_l(j)  
   \right],
\eeq
\beq
\hat H_{\rm QF}(t)
=
\sum_j\hbar\kappa \hat\sigma_+(j)
\sum_m E_m(t)
{\rm e}^{i[\mathbf{k}_m\cdot\mathbf{r}_j-\omega t]}
+
\rm{c.c}.
\eeq
$\hat H_0$ describes the free evolution of the qubit and 
the reservoir systems. 
The index $j$ refers to the $j$th site of qubit 
whose ground and excited states are 
$\vert\downarrow\rangle_j$ and $\vert\uparrow\rangle_j$, 
respectively.  
$\nu_j$ is the energy separation of the $j$th qubit.  
$\hat\sigma_z$ is the Pauli $z$ component operator.  
We assume that the qubit at the $j$th site interacts with 
its own reservoir around the site, 
and 
that the reservoirs at different sites are independent 
but have an identical mode structure $\{\Omega_l\}$ and 
a temperature $T$.  
The index $l$ is the $l$th mode of the reservoir boson with 
the angular frequency $\Omega_l$ 
which is described by the creation and annihilation operators 
$\hat B^\dagger_l(j)$ and $\hat B_l(j)$. 
%
%
$\hat H_{\rm QR}$ describes the coupling 
between the qubit and the reservoir systems 
by the coupling constant $g_l$. 
$\hat H_{\rm QF}(t)$ describes the qubit-external field interaction 
in the rotating wave approximation. 
$\kappa$ is its coupling constant.  
$\hat\sigma_+(j)$ acts as 
$\hat\sigma_+(j)\vert\downarrow\rangle_j=\vert\uparrow\rangle_j$. 
$\mathbf{r}_j$ is the coordinate of the $j$th site of qubit. 
$E_m(t)$, $\mathbf{k}_m$, and $\omega$ 
are the external (classical) field amplitude, 
its wave vector, and angular frequency of the $m$th pulse.

The essential physics of the decoherence in the above model 
is the same as that in the model consisting of 
a single qubit and a bosonic thermal reservoir, 
and has already been clarified by 
extensive theoretical analyses 
\cite{PalmaSuominenEkert96,ViolaLloyd98,Ban98,Uchiyama02}. 
The new feature included in the above model 
is just an array structure of qubits. 
Main concern here is to know 
how the decoherence occuring at a single qubit level 
can be detected in the multiwave mixing signals 
from this array structure.

\begin{figure} 
\begin{center} 
\includegraphics[width=11cm]{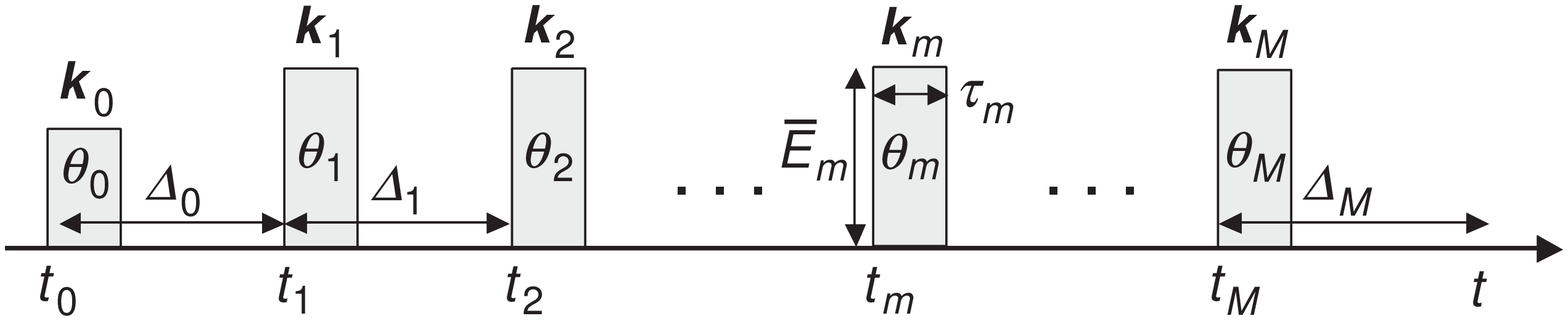} 
\end{center} 
\caption{
\label{PulseSequence} 
Optical pulse sequence. 
}
\end{figure}

The interaction pictured Hamiltonians are given by 
\begin{equation}
\tilde H_{\rm QR}(t)
=
\sum_j\hbar\hat\sigma_z(j)
\sum_l
   \left[
     g_l \hat B^\dagger_l(j) {\rm e}^{i\Omega_l t} 
   + g^\ast_l \hat B_l(j)  {\rm e}^{-i\Omega_l t}
   \right], 
\end{equation}
\beq
\tilde H_{\rm QF}(t)
=
\sum_j\hbar\kappa\hat\sigma_+(j)
\sum_m E_m(t)
{\rm e}^{i[\mathbf{k}_m\cdot\mathbf{r}_j+(\nu_j-\omega) t]}
+
\rm{h.c}. 
\eeq
For simplicity, 
we assume that the optical field in 
$\tilde H_{\rm QF}(t)$ 
are composed of a sequence of optical pulses 
with rectangular temporal shape 
as shown in Fig. \ref{PulseSequence}
which are specified by 
\begin{equation}
E_m(t)
=
\biggl\{
\begin{array}{ll}
{\bar E}_m & (t_m \le t \le t_m + \tau_m) \\
0          & {\rm otherwise}
\end{array}. 
\end{equation}
We further assume that 
the pulse duration $\tau_m$ is so short that the effect of the 
qubit-reservoir interaction $\tilde H_{\rm QR}(t)$ can be 
neglected during the period $[t_m,t_m+\tau_m]$. 
Taking the limit of 
$\tau_m\rightarrow0$ and $\bar E_m\rightarrow\infty$ 
with the finite pulse area 
${\theta_m}=-2i{\bar E}_m \tau_m \kappa$, 
the state evolution under the $m$th optical pulse is described 
by the unitary operator 
\begin{equation}
\tilde U^{(m)}_{\rm QF}
=\prod_j \tilde u^{(m)}_{\mathrm{QF}j}
\label{U_QF2}
\end{equation} 
with 
\begin{equation}
\tilde u^{(m)}_{\mathrm{QF}j}
=
\mathrm{exp}
\left\{
{\theta_m\over2}
\left[
 \hat\sigma_+(j)\mathrm{ e}^{ i\phi^{(m)}_j}
-\hat\sigma_-(j)\mathrm{ e}^{-i\phi^{(m)}_j}
\right]
\right\}, 
\end{equation}
where 
$\phi^{(m)}_j
=\mathbf{k}_m\cdot\mathbf{r}_j + (\nu_j-\omega) t_m$. 
Each qubit evolves according to 
\begin{equation}\label{u_QF1}
\tilde u^{(m)}_{\mathrm{QF}j}\vert\downarrow\rangle_j
=
\mathrm{cos}{\theta_m\over2}\vert\downarrow\rangle_j
+
\mathrm{e}^{i\phi^{(m)}_j}
\mathrm{sin}{\theta_m\over2}\vert\uparrow\rangle_j,
\end{equation}
\begin{equation}\label{u_QF2}
\tilde u^{(m)}_{\mathrm{QF}j}\vert\uparrow\rangle_j
=
\mathrm{cos}{\theta_m\over2}\vert\uparrow\rangle_j
-\mathrm{e}^{-i\phi^{(m)}_j}
\mathrm{sin}{\theta_m\over2}\vert\downarrow\rangle_j.
\end{equation}

The state evolution between the pulses are described by the 
qubit-reservoir interaction  

\noindent
\begin{mathletters}
\begin{eqnarray}
\tilde U^{(m)}_\mathrm{QR}
&=&
T{\rm exp}\left[
                -{i\over\hbar}
                \int_{t_m}^{t_{m+1}}dt'
                \tilde H_{\rm QR}(t')
         \right]
\\
&=&
\prod_j \tilde u^{(m)}_{\mathrm{QR}j},
\label{U_QR}
\end{eqnarray}
\end{mathletters}

\noindent
with 
\begin{equation}
\tilde u^{(m)}_{\mathrm{QR}j}
=
\mathrm{ exp}
\left[
i\Theta(t_{m+1},t_{m})
+
\hat\sigma_z(j)
\sum_l
\left(
 \alpha^{(m)}_l \hat B^\dagger_l(j)
-\alpha^{(m)\ast}_l \hat B_l(j)
\right)
\right],  
\end{equation}
where
\begin{equation}\label{Theta}
\Theta(t_{m+1},t_{m})
=
\sum_l \vert g_l\vert^2
\frac{ \Omega_l(t_{m+1}-t_{m})
      -\mathrm{sin}
       \left[ \Omega_l(t_{m+1}-t_{m}) \right] } 
{\Omega_l^2}, 
\end{equation}
and 
\begin{equation}\label{alpha}
\alpha^{(m)}_l
=
{g_l\over\Omega_l}
\left[
\mathrm{e}^{i\Omega_l t_m}
-
\mathrm{e}^{i\Omega_l t_{m+1}}
\right]. 
\end{equation}
To evaluate the state evolution by 
$\tilde U^{(m)}_\mathrm{QR}$, 
we may use the formula 
\begin{equation}\label{u_QR1}
\tilde u^{(m)}_{\mathrm{QR}j}
\vert\downarrow\rangle_j
\otimes
\vert\Psi\rangle_{\mathrm{R}j}
=
\vert\downarrow\rangle_j
\otimes
\hat D^{(m)}_{j-}
\vert\Psi\rangle_{\mathrm{R}j},
\end{equation}
\begin{equation}\label{u_QR2}
\tilde u^{(m)}_{\mathrm{QR}j}
\vert\uparrow\rangle_j
\otimes
\vert\Psi\rangle_{\mathrm{R}j}
=
\vert\uparrow\rangle_j
\otimes
\hat D^{(m)}_{j+}
\vert\Psi\rangle_{\mathrm{R}j},
\end{equation}
where 
the state 
$\vert\Psi\rangle_{\mathrm{R}j}$ 
represents any pure state component of the $j$th reservoir 
and 
\begin{equation}
\hat D^{(m)}_{j\pm}
\equiv
\prod_l
\hat D_j(\pm \alpha^{(m)}_l) 
\end{equation}
with the displacement operators 
\begin{equation}\label{displacement_op}
\hat D_j(\pm \alpha^{(m)}_l) 
=\mathrm{exp}
\left(
 \pm\alpha^{(m)}_l \hat B^\dagger_l(j)
\mp\alpha^{(m)\ast}_l \hat B_l(j)
\right).
\end{equation}
For the state 
$\vert\Psi\rangle_{\mathrm{R}j}$, 
we will later substitute the Fock state basis 
$\vert n_1...n_l... \rangle_j$ 
which describes the occupation of the reservoir boson modes.

The total evolution operator for the $M$ pulse sequence 
is given by 
%
%
%
\beq
\tilde U(t,t_0)
=
\prod_j \tilde u_j(t,t_0), 
\eeq
with 
\beq
\tilde u_j(t,t_0)
=\tilde u^{(M)}_{\mathrm{QR}j} 
 \tilde u^{(M)}_{\mathrm{QF}j}
\cdot
 \tilde u^{(M-1)}_{\mathrm{QR}j} 
 \tilde u^{(M-1)}_{\mathrm{QF}j}
\cdots
 \tilde u^{(1)}_{\mathrm{QR}j} 
 \tilde u^{(1)}_{\mathrm{QF}j}
 \tilde u^{(0)}_{\mathrm{QR}j} 
 \tilde u^{(0)}_{\mathrm{QF}j}. 
\eeq
As for $\tilde u^{(M)}_{\mathrm{QR}j}$, 
we use 
\begin{equation}\label{alpha2}
\alpha^{(M)}_l
=
{g_l\over\Omega_l}
\left[
\mathrm{e}^{i\Omega_l t_M }
-
\mathrm{e}^{i\Omega_l t} 
\right], 
\end{equation}
i.e., $t_{M+1}$ is understood as $t$, 
instead of the definition of Eq. (\ref{alpha}). 
The final state of the total system is represented by 
\beq
\tilde\rho_{\rm QR}(t)=
\tilde U(t,t_0)
\tilde\rho_{\rm QR}(t_0)
\tilde U(t,t_0)^\dagger, 
\eeq
where we assume the initial state like 
\begin{equation}
\tilde\rho_{\rm QR}(t_0)
=
\prod_j \vert\downarrow\rangle_j\langle\downarrow\vert
\otimes
\prod_l \hat\rho_{{\rm R}j,l}(T)
\end{equation}
with the thermal state 
\begin{eqnarray}
\hat\rho_{{\rm R}j,l}(T)
&=&
(1-{\rm e}^{-\hbar\Omega_l/k_{\rm B}T})
{\rm e}^{-\hbar\Omega_l \hat B^\dagger_l (j) 
                        \hat B_l (j) /k_{\rm B}T}, 
\\
&=&
\sum_{n_l(j)=0}^\infty 
\frac{\bar n_l^{n_l(j)}}
     {(1+\bar n_l)^{n_l(j)+1}}
     \vert n_l(j)\rangle_{\mathrm{R}j}\langle n_l(j)\vert, 
\end{eqnarray}
where 
\beq
\bar n_l=1/
\left( 
       \mathrm{e}^{\hbar\Omega_l/k_{\rm B}T}-1
\right).
\eeq

By using the formula Eqs. (\ref{u_QR1}) and (\ref{u_QR2}), 
the final state can be represented as 
\beq
\tilde\rho_{\rm QR}(t)
=
\prod_j
\left[
\sum_{\vec n(j)} 
\left(
\prod_{l=1}^\infty
\frac{\bar n_l^{n_l(j)}}
     {(1+\bar n_l)^{n_l(j)+1}}
\right)
\vert \tilde\rho_{\vec n(j)}(t) 
\rangle_{{\rm QR}j}
\langle \tilde\rho_{\vec n(j)}(t) \vert
\right],
\eeq
where 
\beq\label{rho_QR}
\vert \tilde\rho_{\vec n(j)}(t) \rangle_{{\rm QR}j}
\equiv
\tilde u_j(t,t_0)
  \vert\downarrow\rangle_j
  \otimes
  \vert\vec n(j)\rangle_{\mathrm{R}j}, 
\eeq
and 
$\vec n(j) \equiv(n_0(j),...,n_l(j),...)$ 
denoting the boson number occupation. 
To describe the state of Eq. (\ref{rho_QR}), 
we introduce the two operators 
$\hat F^{(M)}_j$ and $\hat G^{(M)}_j$ by 
\beq
\vert\tilde\rho_{\vec n(j)}(t)\rangle_{{\mathrm QR}j}
\equiv
(-1)^{M\over2}
\biggl(
\vert\downarrow\rangle_j\otimes \hat G^{(M)}_j
+
\vert\uparrow\rangle_j\otimes \hat F^{(M)}_j
\biggr)
\vert\vec n(j)\rangle_{{\mathrm R}j},
\eeq
for $M=$even, 
and  
\beq
\vert\tilde\rho_{\vec n(j)}(t)\rangle_{{\mathrm QR}j}
\equiv
(-1)^{{M-1}\over2}
\biggl(
-\vert\downarrow\rangle_j\otimes \hat F^{(M)}_j
+
\vert\uparrow\rangle_j\otimes \hat G^{(M)}_j
\biggr)
\vert\vec n(j)\rangle_{{\mathrm R}j},
\eeq
for $M=$odd.

\section{State evolution under $\pi$ pulses}
\label{State evolution under pi pulses}

\subsection{Formulation}
\label{Pi:Formulation}

In this section we consider the pulse sequence 
consisting of 
the 0th exciting pulse with the area $\theta_0=\pi/2$ 
at time $t_0$, 
and 
the subsequent $M$ $\pi$-pulses, 
i.e., 
$\theta_m=\pi$ $(m=1,...,M)$. 
By using Eqs. 
(\ref{u_QF1}), (\ref{u_QF2}), (\ref{u_QR1}), and (\ref{u_QR2}), 
we obtain 
\begin{equation}
\hat F^{(M)}_j
=
{1\over{\sqrt{2}}}
\mathrm{e}^{i(\Phi^{(M)}_j + \phi^{(0)}_j)}
\hat D^{(M)}_{j(-1)^{M}} 
\cdots 
\hat D^{(1)}_{j-} 
\hat D^{(0)}_{j+}
\end{equation}
and 
\begin{equation}
\hat G^{(M)}_j
=
{1\over{\sqrt{2}}}
\mathrm{e}^{-i \Phi^{(M)}_j }
\hat D^{(M)}_{j(-1)^{M+1}} 
\cdots 
\hat D^{(1)}_{j+} 
\hat D^{(0)}_{j-}
\end{equation}
where 
\begin{equation}\label{Phi_j}
\Phi^{(M)}_j
=
\sum_{m=1}^M (-1)^m \phi^{(m)}_j. 
\end{equation}

The decoherence property of the qubit system is measured 
by observing the intensity of the radiation from 
the macroscopic polarization. 
The polarization operators in the interaction picture 
are defined by 
\beq
\tilde S_{\pm}(\mathbf{q},t)
\equiv
\sum_j \hat \sigma_{\pm}(j)
\mathrm{e}^{i\mathbf{q}\cdot\mathbf{r}_j \pm \nu_j t}. 
\eeq 
The macroscopic polarization is then evaluated by 
\begin{eqnarray}\label{Polarization_Pi1}
P(\mathbf{q}, t) 
&=&
{\rm Tr}
\left[
      \sum_j \tilde S_+(\mathbf{q},t)
      \tilde\rho_{\rm QR}(t)
\right]
\mathrm{e}^{-i\omega t},
\nonumber\\
&=&
\sum_j {\rm e}^{i[\mathbf{q}\cdot\mathbf{r}_j+(\nu_j-\omega)t]} 
{\rm Tr}
\left[
      \tilde\sigma_+(j) 
      \tilde\rho_{\rm QR}(t)
\right],
\nonumber\\
&=&
\Biggl\{
\begin{array}{lll}
{}&
\displaystyle\sum_j 
{\rm e}^{i[\mathbf{q}\cdot\mathbf{r}_j+(\nu_j-\omega)t]} 
{\rm Tr}_{\rm R}
\left[
      \hat F^{(M)\dagger}_j \hat G^{(M)}_j \prod_l 
      \hat \rho_{{\rm R}j,l}(T)
\right],   
& (M={\rm even}), 
\\
{}&{}&{}
\\
-&
\displaystyle\sum_j 
{\rm e}^{i[\mathbf{q}\cdot\mathbf{r}_j+(\nu_j-\omega)t]}     
{\rm Tr}_{\rm R}
\left[
      \hat G^{(M)\dagger}_j \hat F^{(M)}_j \prod_l 
      \hat \rho_{{\rm R}j,l}(T)
\right],
& (M={\rm odd}),
\end{array} 
\nonumber\\
&=&
\frac{(-1)^M}{2} 
\sum_j {\rm exp}
i\left[
       \mathbf{q}\cdot\mathbf{r}_j + (\nu_j-\omega)t 
      +(-1)^{M+1}\left( 2\Phi_j^{(M)}+\phi_j^{(0)} \right)
\right]
\nonumber\\
&{}&\times
{\rm Tr}_{\rm R}
\left[
\prod_l 
\hat D_j \left( 
             (-1)^{M+1}2\sum_{m=0}^M (-1)^m \alpha_l^{(m)} 
       \right)
\hat \rho_{{\rm R}j,l}(T)
\right], 
\end{eqnarray}
Let us introduce the decoherence exponent by 
\beq
{\rm e}^{-\Gamma(t)}
\equiv
{\rm Tr}_{\rm R}
\left[
\prod_l 
\hat D_j \left( 
             (-1)^{M+1}2\sum_{m=0}^M (-1)^m \alpha_l^{(m)} 
       \right)
\hat \rho_{{\rm R}j,l}(T)
\right]. 
\eeq
The decoherence exponent, 
which is now {\it independent} of $j$, 
is expressed as 
\cite{PalmaSuominenEkert96,ViolaLloyd98} 
\begin{eqnarray}\label{DecoherenceExponent}
\Gamma(t)
&=&
{1\over2}
\sum_l 
\Biggl\vert 2 
      \sum_{m=0}^M (-1)^m \alpha_l^{(m)} 
\Biggr\vert^2
\mathrm{coth}
\left(  \frac{\hbar\Omega_l}{2k_\mathrm{B}T}  \right)
\nonumber\\
&=&
2\sum_l \frac{\vert g_l\vert^2}{\Omega^2_l}
f(\Omega_l,t)
\mathrm{coth}
\left(  \frac{\hbar\Omega_l}{2k_\mathrm{B}T}  \right),
\nonumber\\
&=& 
2\int_{0}^\infty d\Omega
I(\Omega)
\frac{f(\Omega,t)}{\Omega^2}
\mathrm{coth}
\left(  \frac{\hbar\Omega}{2k_\mathrm{B}T}  \right),
\end{eqnarray}
where  
\beq
I(\Omega)
=
\sum_l \delta(\Omega-\Omega_l) \vert g_l\vert^2, 
\eeq
and 
\beq
f(\Omega,t)
=
\left\vert
\sum_{m=0}^{M-1}
(-1)^m 
({\rm e}^{i\Omega t_m}-{\rm e}^{i\Omega t_{m+1}}) 
+
(-1)^M 
({\rm e}^{i\Omega t_M}-{\rm e}^{i\Omega t}) 
\right\vert^2.  
\eeq

By substituting the definition of 
$\Phi_j^{(M)}$ (Eq. (\ref{Phi_j})) 
and 
$\phi^{(m)}_j
=\mathbf{k}_m\cdot\mathbf{r}_j + (\nu_j-\omega) t_m$ 
into Eq. (\ref{Polarization_Pi1}), 
we have 
\beq\label{Polarization_Pi2}
P(\mathbf{q}, t) 
=
\frac{(-1)^M}{2}
{\rm e}^{-\Gamma(t)}
\sum_j 
{\rm exp}
i\left[
         \left(
               \mathbf{q}-(-1)^M \mathbf{K}^{(M)}
         \right)
               \cdot\mathbf{r}_j 
       + (\nu_j-\omega) 
         \left( \Delta_M + 
                (-1)^M \sum_{m=0}^{M-1} (-1)^m \Delta_m
         \right)
\right], 
\eeq
where 
\begin{equation}\label{PhaseMatching Direction}
\mathbf{K}^{(M)}
=
2\sum_{m=1}^M (-1)^m \mathbf{k}_m + \mathbf{k}_0, 
\end{equation}
$\Delta_m\equiv t_{m+1}-t_{m}$, 
and 
$\Delta_M\equiv t - t_{M}$. 
Eq. (\ref{Polarization_Pi2}) shows that 
the radiation signal is observed in the phase matching direction 
$\mathbf{q}=(-1)^M \mathbf{K}^{(M)}$.

\subsection{Homogeneous case}
\label{Pi:Homogeneous case}

We first consider the homogeneous case of $\nu=\nu_i$ for all $i$. 
This case was already studied by Aihara 
as the photon echo in the four wave mixing (two pulse) scheme 
\cite{Aihara80,Aihara82}. 
He pointed out that the interaction between the qubit and 
the reservoir bosons induces the slow frequency modulation 
on the qubit in the short time region 
where the reservoir keeps the phase memory, 
and plays the similar role to the inhomogeneous broadening. 
In fact he predicted that the photon echo must be observed
even in the homogeneous case.
The BB control proposed by Ban 
\cite{Ban98}, 
and Viola and Lloyd 
\cite{ViolaLloyd98}
can be regarded as a natural extension of this principle 
to the multi-pulse scheme. 
It is well studied how to design a pulse sequence 
for effective suppression of decoherence 
\cite{Ban98,ViolaLloyd98,Uchiyama02}. 
The decoherence can be made worse 
by applying a pulse sequence, 
if the pulse intervals are not taken short enough.

\begin{figure} 
\begin{center} 
\includegraphics[width=17cm]{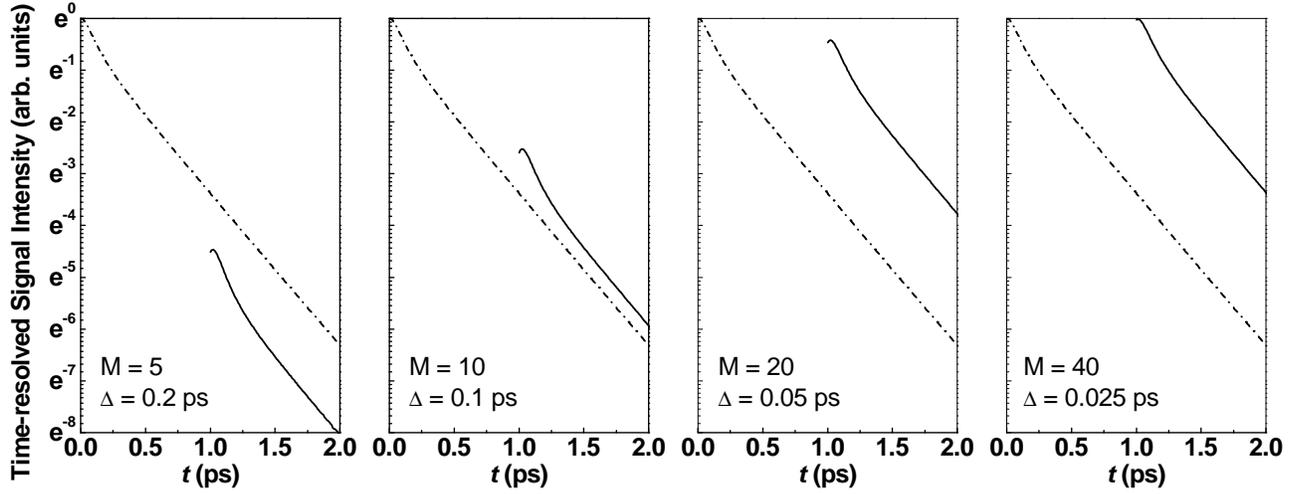} 
\end{center} 
\caption{
\label{pi40ohmic10kwc8a1} 
The time-resolved multiwave mixing signal intensities 
under the multi $\pi$ pulse 
irradiation (solid lines) compared with the ones of 
the free induction decay (one dotted lines) 
in the case of the ohmic reservoir model. 
The parameters are 
$\Omega_{\rm c}$=8\,meV, 
$\alpha=0.1$ and $T$=10\,K. 
$M$ is the number of pulses and 
$\Delta$ is the pulse interval. 
}
\end{figure}
\begin{figure} 
\begin{center} 
\includegraphics[width=17cm]{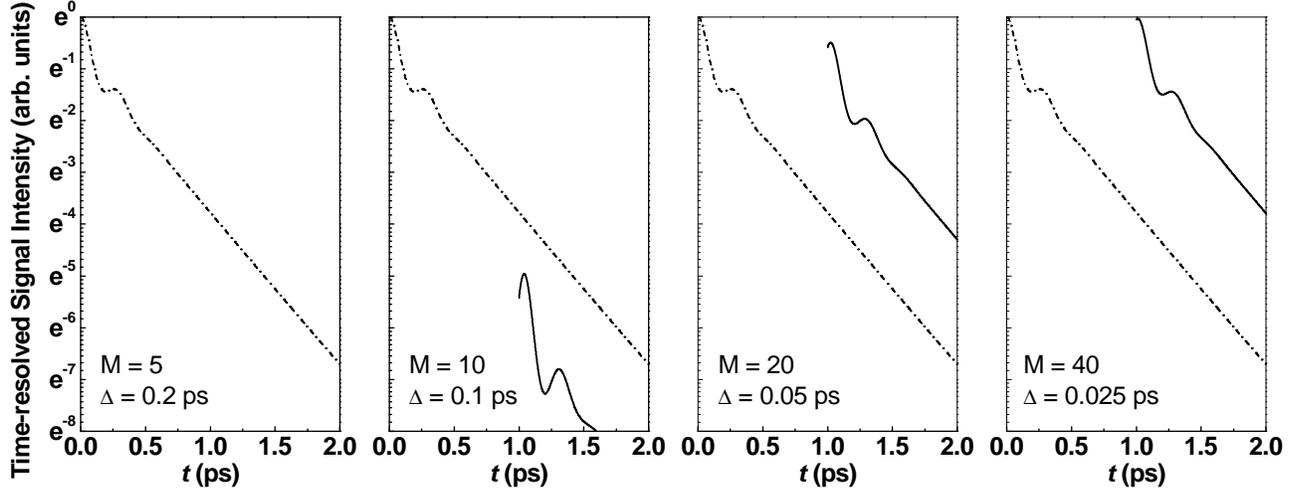} 
\end{center} 
\caption{
\label{pi40gauss10kwc8a1wp13gp4ap05} 
The time-resolved multiwave mixing signal intensities 
under the multi $\pi$ pulse 
irradiation (solid lines) compared with the ones of 
the free induction decay (one dotted lines) 
in the case of the Gaussian-ohmic reservoir model. 
The parameters are 
$\Omega_{\rm p}$=13\,meV, 
$\gamma_{\rm p}$=4\,meV,
$\alpha_{\rm p}=0.05$, 
$\Omega_{\rm c}$=8\,meV, 
$\alpha=0.1$, 
and $T$=10\,K. 
$M$ is the number of pulses and 
$\Delta$ is the pulse interval. 
}
\end{figure}

Fig. \ref{pi40ohmic10kwc8a1} shows numerical results 
of the decoherence under an $M$ $\pi$ pulse sequence 
with an interval $\Delta$. 
The solid lines represent the time-resolved multiwave 
mixing signal intensities of 
the multiwave mixing signals, 
$I_{\mathbf{q}}(t)=\vert P(\mathbf{q},t) \vert^2$, 
in the phase matching direction 
$\mathbf{q}=(-1)^M \mathbf{K}^{(M)}$ 
(Eq. (\ref{PhaseMatching Direction})), 
while 
the one dotted lines represent the ones of 
the free induction decay signals 
in the direction $\mathbf{k}_0$. 
We have assumed the ohmic reservoir model characterized 
by the spectral density 
\begin{equation}
I(\Omega)
=
\alpha\Omega{\rm exp}(-\frac{\Omega}{\Omega_{\rm c}}), 
\end{equation}
where 
$\Omega_{\rm c}$ is the cutoff frequency and 
$\alpha$ is the dimensionless coupling constant. 
As seen, 
for a pulse interval of $\Delta=0.2$\,ps, 
the decoherece is not suppressed but accelerated. 
In Fig. \ref{pi40gauss10kwc8a1wp13gp4ap05}, 
we have shown the same kind of results 
in the case of the Gaussian-ohmic reservoir model, 
which 
has not only the ohmic character with a wider spectrum 
but also 
the Gaussian character with a sharper spectrum 
around a characteristic frequency $\Omega_{\rm p}$, 
such as a specific phonon mode. 
The decay profile exhibits the oscillation characterized 
by the frequency $\Omega_{\rm p}$. 
For shorter pulse intervals, 
the decoherence is accelerated by adding $\pi$ pulses. 
\beq
I(\Omega)
=
\alpha\Omega{\rm exp}(-\frac{\Omega}{\Omega_{\rm c}})
+
\frac{\alpha_{\rm p}\Omega^2}
     {\sqrt{\pi}\gamma_\mathrm{p}}
{\rm exp}
\left[
       -\frac{(\Omega-\Omega_\mathrm{p})^2}
             {\gamma_\mathrm{p}^2}
\right]. 
\eeq

Although this decoherence acceleration phenomenon 
is essential to understand the decoherence, 
its interpretation given so far is unsatisfactory. 
Here we revisit this point by considering a simplified model 
\beq\label{Hamiltonian_QR_single_mode}
\hat H_{{\rm QR}_{\rm S}}(t)
=
\hbar\hat\sigma_z 
(  g_\mathrm{p} \hat B^\dagger + g^\ast_\mathrm{p} \hat B  ), 
\eeq 
i.e., a signal qubit interacts with a single mode field of
the reservoir boson oscillator
with the angular frequency $\Omega_\mathrm{p}$. 
Under this Hamiltonian, 
the qubit couples with the reservoir as 
\begin{eqnarray}
&&
(\vert\uparrow\rangle+\vert\downarrow\rangle)
  \otimes
  \vert n\rangle_{\rm R}
\nonumber\\
&\rightarrow&
\left(  
   \vert\uparrow\rangle
     \otimes 
     \hat D[ \alpha(t-t_0)] 
 + \vert\downarrow\rangle
     \otimes 
     \hat D[-\alpha(t-t_0)] 
\right)
     \vert n\rangle_{\rm R}, 
\end{eqnarray}
where 
$\vert n\rangle_{\rm R}$ is the {\it n} boson state component, 
and 
\begin{equation}
\hat D\left(\pm\alpha(t-t_0)\right)
=
\mathrm{exp}\left[\pm\alpha(t-t_0)
\hat B^\dagger\mp\alpha(t-t_0)^\ast \hat B\right],
\end{equation}
with
\beq
\alpha(t-t_0)=\frac{g_\mathrm{p}}{\Omega_\mathrm{p}}
(  {\rm e}^{i\Omega_\mathrm{p} t_0}
 - {\rm e}^{i\Omega_\mathrm{p} t}). 
\eeq
The free induction decay is characterized by 
the decoherence exponent 
\beq
\Gamma_+(t)
=
2 \frac{\vert g_\mathrm{p}\vert^2}{\Omega^2_\mathrm{p}}
f_+(t)
\mathrm{coth}
\left(  
      \frac{\hbar\Omega_\mathrm{p}}{2k_\mathrm{B}T}  
\right)
\equiv
g' f_+(t),  
\eeq
where 
\begin{eqnarray}\label{Gamma_p}
f_+(t)
&=&
       \vert
           {\rm e}^{i\Omega_\mathrm{p} t_0}
         - {\rm e}^{i\Omega_\mathrm{p} t}
       \vert^2
\nonumber\\
&=&
2  \Biggl\{
         1-\mathrm{cos}  \Bigl[ \Omega_\mathrm{p}(t-t_0) \Bigr]  
   \Biggr\}. 
\end{eqnarray}
Thus $\Gamma_+(t)$ oscillates in time and 
the qubit is recohered, i.e.,
disentangled from the reservoir, at 
$t-t_0=(2\pi/\Omega_\mathrm{p})\times$(integer). 
On the other hand, 
when a $\pi$ pulse is applied at a time $t_1$, 
the decoherence exponent is given by 
\beq
\Gamma_-(t) \equiv g' f_-(t)
\eeq
with 
\begin{eqnarray}
f_-(t)
&=&    \vert
           {\rm e}^{i\Omega_\mathrm{p} t_0}
         - {\rm e}^{i\Omega_\mathrm{p} t_1} 
         -({\rm e}^{i\Omega_\mathrm{p} t_1}
         - {\rm e}^{i\Omega_\mathrm{p} t})
       \vert^2, 
\nonumber\\
&=&
  2\bigl[3-2\mathrm{cos}(\Omega_\mathrm{p}\Delta_0)\bigr]
\nonumber\\
&-&
2\sqrt{5-4\mathrm{cos}(\Omega_\mathrm{p}\Delta_0)}
\mathrm{cos}[\Omega_\mathrm{p}(t-t_0)+\varphi],
\end{eqnarray}
where 
\beq
\mathrm{tan}\varphi
=
\frac{2\mathrm{cos}\Omega_\mathrm{p}\Delta_0}
 {1-2 \mathrm{cos}\Omega_\mathrm{p}\Delta_0}.
\eeq
For small enough $\Delta_0$ $(\equiv t_1-t_0)$, 
\beq
\Gamma_-(t)
\sim 
2 g' 
\left\{
   1-\mathrm{cos}
  \left[   \Omega_\mathrm{p} (t-t_0-2\Delta_0)  \right]
\right\}. 
\eeq
This should be compared with $\Gamma_+(t)$. 
We can see that the dephasing exponent $\Gamma_-(t)$ is 
the one shifted in time by $2\Delta_0$ from $\Gamma_+(t)$. 
Namely, by applying the $\pi$ pulse
soon after the state 
$\vert\uparrow\rangle+\vert\downarrow\rangle$
is prepared, 
the decoherence starts to take place 
at the time $2\Delta_0$ later 
compared to the free induction decay 
without any $\pi$ pulse. 
For larger $\Delta_0$,  however,  this is not true. 
As an extreme case, 
we take $\Omega_\mathrm{p}\Delta_0=\pi$.
Then 
\beq
\Gamma_-(t)
=g'
\left\{
   6\mathrm{cos}
      \left[   \Omega_\mathrm{p} (t-t_0)   \right]
   +10
\right\}
\ge 
\Gamma_+(t). 
\eeq
In this case, 
the $\pi$ pulse plays a role to accelerate the decoherence. 
The reason is clear from the structure of $f_-(t)$. 
When the $\pi$ pulse is applied after 
$t_1-t_0>\pi/(2\Omega_\mathrm{p})$,  
the two successive displacement operations in the state component 
\beq
\left(  
   \vert\uparrow\rangle
     \otimes 
     \hat D[ \alpha(t-t_1) ] \hat D[-\alpha(t_1-t_0) ] 
 - \vert\downarrow\rangle
     \otimes 
     \hat D[-\alpha(t-t_1) ] \hat D[ \alpha(t_1-t_0) ]
\right)
     \vert n\rangle_{\rm R}, 
\eeq
act as 
{\it in-phase} in the sense that the displacements 
by the amount of $\alpha(t-t_1)$ and $-\alpha(t_1-t_0)$ 
do not cancel out 
but are added in the same sign, 
leading to large degree of entanglement between the qubit 
and the reservoir. 
On the other hand,  
when no $\pi$ pulse is applied, 
the state evolves as 
\beq
\left(  
   \vert\uparrow\rangle
     \otimes 
     \hat D[ \alpha(t-t_1) ] \hat D[ \alpha(t_1-t_0) ] 
 + \vert\downarrow\rangle
     \otimes 
     \hat D[-\alpha(t-t_1) ] \hat D[-\alpha(t_1-t_0) ]
\right)
     \vert n\rangle_{\rm R}, 
\eeq 
where the two displacement operations act as 
{\it out-of-phase}, 
i.e., 
$\alpha(t-t_1)$ and $\alpha(t_1-t_0)$ nearly cancel out 
for $t_1-t_0>\pi/(2\Omega_\mathrm{p})$, 
leading to the recoherence of the qubit superposition state.

In a realistic thermal reservoir with many modes, 
the above feature is more or less smeared out, 
but can still be identified at low temperatures. 
Roughly speaking, 
we had better to take the $\pi$ pulse interval as 
\begin{equation}
\label{stabilization condition}
\Delta<\pi/(2\Omega_\mathrm{th}),  
\end{equation}
where $\Omega_\mathrm{th}$ is the largest characteristic 
frequency of the thermalized reservoir boson spectrum 
$
I(\Omega)
{\rm coth}(\frac{\hbar\Omega}{2k_{\rm B}T}). 
$
At low temperatures, 
$\Omega_\mathrm{th}$ is roughly 
$\Omega_\mathrm{c}$ in the ohmic reservoir model 
(Fig. \ref{pi40ohmic10kwc8a1}), 
or  
$\Omega_\mathrm{p}$ in the Gaussian-ohmic reservoir model 
(Fig. \ref{pi40gauss10kwc8a1wp13gp4ap05}).

\subsection{Inhomogeneous case}
\label{Pi:Inhomogeneous case}

\begin{figure} 
\begin{center} 
\includegraphics[width=17cm]{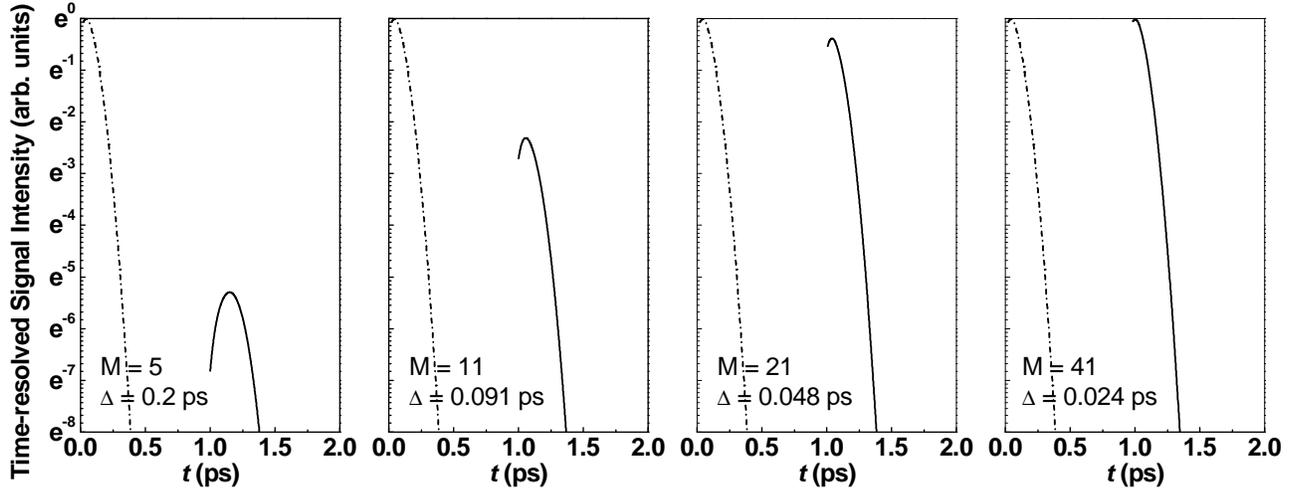} 
\end{center} 
\caption{
\label{pi41ohmicinhm10kwc8a1} 
The time-resolved multiwave mixing signal intensities 
under the multi $\pi$ pulse 
irradiation (solid lines) compared with the ones of 
the single $\pi$ pulse, i.e., $M=1$ (one dotted lines) 
in the case of the ohmic reservoir model 
when the inhomogeneous broadening effect is taken into account 
by setting 
$\delta_{\rm B}$=5meV. 
The other parameters are the same as in 
Fig. \ref{pi40ohmic10kwc8a1}. 
}
\end{figure}
\begin{figure} 
\begin{center} 
\includegraphics[width=17cm]{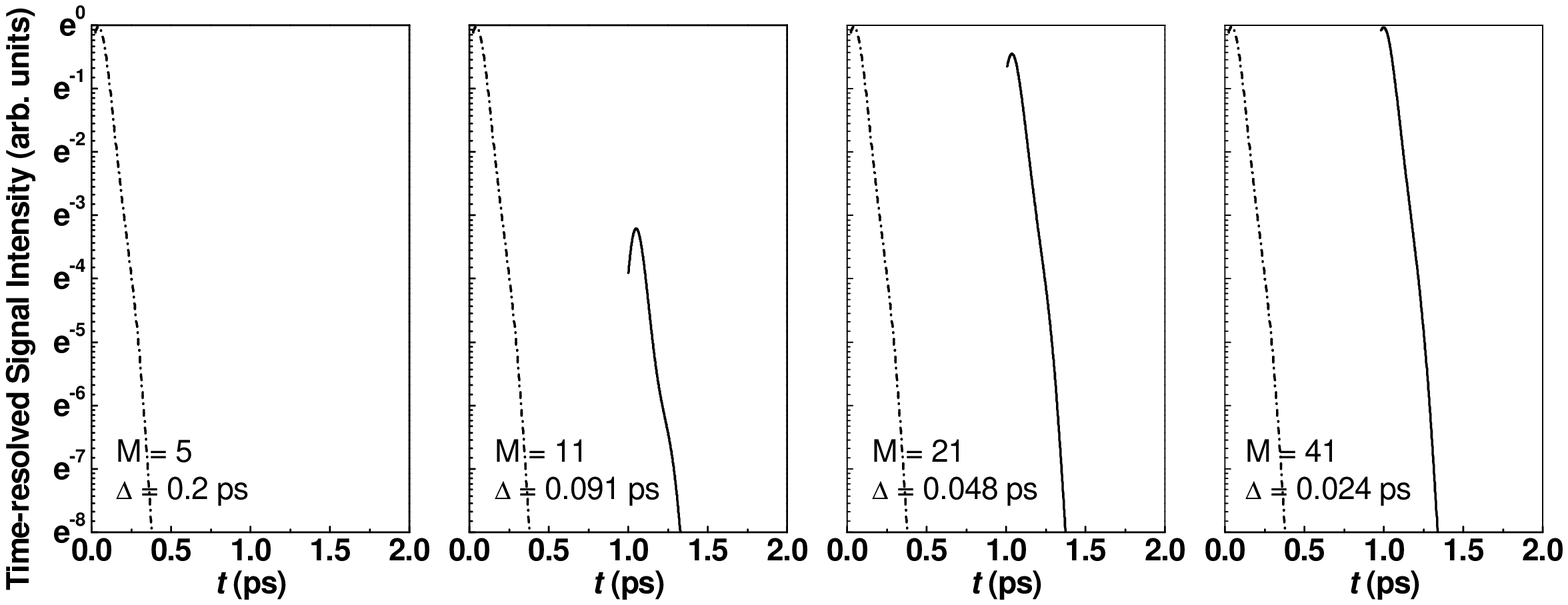} 
\end{center} 
\caption{
\label{pi41gaussinhm10kwc8a1wp13gp4ap05} 
The time-resolved multiwave mixing signal intensities 
under the multi $\pi$ pulse 
irradiation (solid lines) compared with the ones of 
the single $\pi$ pulse, i.e., $M=1$ (one dotted lines) 
in the case of the Gaussian-ohmic resevoir model 
when the inhomogeneous broadening effect is taken into account 
by setting 
$\delta_{\rm B}$=5meV. 
The other parameters are the same as in 
Fig. \ref{pi40gauss10kwc8a1wp13gp4ap05}. 
}
\end{figure}

In this subsection we take into account 
the inhomogeneous broadening effect 
in a bulk ensemble of qubits. 
Now we assume that the inhomogeneous broadening of $\nu_j$ obeys 
the Gaussian distribution around the angular frequency $\omega$ 
of the external field 
\beq\label{Inhomo_broadening_distribution}
p(\nu-\omega)
=
\frac{1}{\sqrt{2\pi\delta_\mathrm{B}^2}}
\mathrm{exp}
\left[
-\frac{(\nu-\omega)^2}{2\delta_\mathrm{B}^2}
\right], 
\eeq
which leads us to 
\begin{eqnarray}
&{}&\sum_j 
{\rm exp}
\left[ i (\nu_j-\omega) 
         \left( \Delta_M + 
                (-1)^M \sum_{m=0}^{M-1} (-1)^m \Delta_m
         \right)
\right]
\nonumber\\
&{}&=
\int_{-\infty}^\infty d\nu 
p(\nu-\omega)
{\rm exp}
\left[ i (\nu-\omega) 
         \left( \Delta_M+(-1)^M 
                \sum_{m=0}^{M-1} (-1)^m \Delta_m
         \right)
\right]
\nonumber\\
&{}&=
{\rm exp}
\left[  
      -{1\over2}
        \left( \Delta_M+(-1)^M 
                \sum_{m=0}^{M-1} (-1)^m \Delta_m
        \right)
        \delta_\mathrm{B}^2
\right]. 
\end{eqnarray}
The time-resolved multiwave mixing signal intensity 
is finally obtained as 
\begin{eqnarray}\label{RadiationSignalIntensity_Pi_inhomo}
I_{(-1)^M \mathbf{K}^{(M)}}(t)
&=&
\Bigl\vert 
    P \bigl( (-1)^M \mathbf{K}^{(M)}, t \bigr) 
\Bigr\vert^2
\nonumber\\
&=&
{1\over4}
{\rm exp}
\left[ 
      -2\Gamma(t) 
      -\left( \Delta_M +(-1)^M 
                \sum_{m=0}^{M-1} (-1)^m \Delta_m
       \right)
       \delta_\mathrm{B}^2
\right]. 
\end{eqnarray}
The decoherence property is evaluated 
by measuring the echo signal.  
In order to see the echo signal, we consider a pulse sequence 
consisting of odd number of $M$ $\pi$ pulses such as 
$\Delta_0=\Delta_1=...=\Delta_{M-2}\equiv \Delta$, 
and the last interval 
$\Delta_{M-1}$ $(=t_M-t_{M-1})$ being varied. 
The echo signal is then expected around $t=t_M+\Delta_{M-1}$ 
as seen from Eq. (\ref{RadiationSignalIntensity_Pi_inhomo}).

Figs. \ref{pi41ohmicinhm10kwc8a1}
and \ref{pi41gaussinhm10kwc8a1wp13gp4ap05} 
shows 
how the results of Figs. 
\ref{pi40ohmic10kwc8a1}
and \ref{pi40gauss10kwc8a1wp13gp4ap05} 
are modified, respectively, 
when the inhomogeneous broadening effect is taken into account. 
We assume $\delta_{\rm B}$=5meV. 
We see essentially the same behaviors as in 
Figs. \ref{pi40ohmic10kwc8a1}
and \ref{pi40gauss10kwc8a1wp13gp4ap05} 
except that the echo signals are more sharpened 
by the fast dephasing due to the inhomogeneity. 
The one dotted lines correspond to the case of 
the single $\pi$ pulse ($M=1$), i.e., 
an ordinary four wave mixing scheme for photon echo experiment.

\section{State evolution under weak pulses}
\label{State evolution under weak pulses}

\subsection{Formulation}
\label{Weak:Formulation}

In an actual optical experiment, it is often the case 
that the maximum laser power falls short of   
the ideal $\pi$ pulse and then the decoherence suppression 
does not work perfectly. In the qubit array system, 
only some fraction of qubits can be suppressed. 
In fact, the intensity of the multiwave mixing signal 
observed in the direction 
$(-1)^M \mathbf{K}^{(M)}$ will be reduced by a factor 
${\rm sin}^4{\theta_1\over2}
\cdots
{\rm sin}^4{\theta_M\over2}$. 
In other words at least this fraction of the qubits can feel 
the ideal $\pi$ pulses, 
and their decoherence will be suppressed. 
From this point of view, 
we consider how one can demonstrate experimentally 
the principle of the BB control.

In the case of weak pulses, 
the analysis is no more straightforward. 
We have the complicated diffraction patterns 
in various directions. 
For simplicity, let us consider the three-pulse case 
consisting the 0th exciting pulse, the 1st and 2nd 
controlling pulses. 
The $\hat F^{(M)}_j$ and $\hat G^{(M)}_j$ operators are given by 
\begin{eqnarray}
\hat F^{(2)\dagger}_j
&=&
  \mathrm{e}^{-i\phi^{(0)}_j}
  \mathrm{sin}{\theta_0\over2}
  \mathrm{cos}{\theta_1\over2}
  \mathrm{cos}{\theta_2\over2}
  \hat D^{(2)}_{j-} \hat D^{(1)}_{j-} \hat D^{(0)}_{j-}
  \mathrm{e}^{i[- \alpha_+^{(210)} - \alpha^{(10)} ]}
\nonumber\\
&+&
  \mathrm{e}^{-i\phi^{(1)}_j}
  \mathrm{cos}{\theta_0\over2}
  \mathrm{sin}{\theta_1\over2}
  \mathrm{cos}{\theta_2\over2}
  \hat D^{(2)}_{j-} \hat D^{(1)}_{j-} \hat D^{(0)}_{j+}
  \mathrm{e}^{i[- \alpha_-^{(210)} + \alpha^{(10)} ]}
\nonumber\\
&+&
  \mathrm{e}^{-i\phi^{(2)}_j}
  \mathrm{cos}{\theta_0\over2}
  \mathrm{cos}{\theta_1\over2}
  \mathrm{sin}{\theta_2\over2}
  \hat D^{(2)}_{j-} \hat D^{(1)}_{j+} \hat D^{(0)}_{j+}
  \mathrm{e}^{i[ \alpha_+^{(210)} - \alpha^{(10)} ]},
\nonumber\\
&+&
  \mathrm{e}^{i[-\phi^{(0)}_j+\phi^{(1)}_j-\phi^{(2)}_j]}
  \mathrm{sin}{\theta_0\over2}
  \mathrm{sin}{\theta_1\over2}
  \mathrm{sin}{\theta_2\over2}
  \hat D^{(2)}_{j-} \hat D^{(1)}_{j+} \hat D^{(0)}_{j-}
  \mathrm{e}^{i[ \alpha_-^{(210)} + \alpha^{(10)} ]},
\end{eqnarray}
and 
\begin{eqnarray}
\hat G^{(2)}_j
&=&
  \mathrm{cos}{\theta_0\over2}
  \mathrm{cos}{\theta_1\over2}
  \mathrm{cos}{\theta_2\over2}
  \hat D^{(2)}_{j-} \hat D^{(1)}_{j-} \hat D^{(0)}_{j-}
  \mathrm{e}^{i[  \alpha_+^{(210)} + \alpha^{(10)} ]}
\nonumber\\
&-&
  \mathrm{e}^{i[\phi^{(0)}_j - \phi^{(1)}_j]}
  \mathrm{sin}{\theta_0\over2}
  \mathrm{sin}{\theta_1\over2}
  \mathrm{cos}{\theta_2\over2}
  \hat D^{(2)}_{j-} \hat D^{(1)}_{j-} \hat D^{(0)}_{j+}
  \mathrm{e}^{i[  \alpha_-^{(210)} - \alpha^{(10)} ]}
\nonumber\\
&-&
  \mathrm{e}^{i[\phi^{(1)}_j - \phi^{(2)}_j]}
  \mathrm{cos}{\theta_0\over2}
  \mathrm{sin}{\theta_1\over2}
  \mathrm{sin}{\theta_2\over2}
  \hat D^{(2)}_{j-} \hat D^{(1)}_{j+} \hat D^{(0)}_{j-}
  \mathrm{e}^{i[- \alpha_-^{(210)} - \alpha^{(10)} ]},
\nonumber\\
&-&
  \mathrm{e}^{i[  \phi^{(0)}_j - \phi^{(2)}_j]}
  \mathrm{sin}{\theta_0\over2}
  \mathrm{cos}{\theta_1\over2}
  \mathrm{sin}{\theta_2\over2}
  \hat D^{(2)}_{j-} \hat D^{(1)}_{j+} \hat D^{(0)}_{j+}
  \mathrm{e}^{i[- \alpha_+^{(210)} + \alpha^{(10)} ]},
\end{eqnarray}
where 
\beq
 \alpha_\pm^{(210)}
 \equiv
 \sum_l \mathrm{Im}
   [ \alpha_l^{(2)} 
     (\alpha_l^{(1)} \pm \alpha_l^{(0)})^\ast ],
\eeq
and
\beq
 \alpha_\pm^{(10)}
 \equiv
 \sum_l \mathrm{Im}
   [ \alpha_l^{(1)} \alpha_l^{(0)\ast} ]. 
\eeq
The polarization is then given by 
\begin{eqnarray}\label{Polarization_weak1}
P(\mathbf{q}, t) 
&=&
  \sum_j 
  {\rm e}^{i[\mathbf{q}\cdot\mathbf{r}_j+(\nu_j-\omega)t]} 
  {\rm Tr}_{\rm R}
  \left[
      \hat F^{(2)\dagger}_j \hat G^{(2)}_j \prod_l 
      \hat \rho_{{\rm R}j,l}(T)
  \right],   
\nonumber\\
&=&
  \sum_j
  {\rm e}^{i[ (\mathbf{q}-\mathbf{k}_0)\cdot\mathbf{r}_j 
             +(\nu_j-\omega)(t-t_2 + \Delta_1+\Delta_0)]}
  {\rm e}^{-\Gamma_{---}(t)}
  {1\over2}
  \mathrm{sin}{\theta_0}
  \mathrm{cos}^2{\theta_1\over2}
  \mathrm{cos}^2{\theta_2\over2}
\nonumber\\
&+&
  \sum_j
  {\rm e}^{i[ (\mathbf{q}-\mathbf{k}_1)\cdot\mathbf{r}_j 
             +(\nu_j-\omega)(t-t_2 + \Delta_1)]}
  {\rm e}^{-\Gamma_{--0}(t)}
  {1\over2}
  \left(
     {\rm e}^{ i\gamma_{--0}(t)}\mathrm{cos}^2{\theta_0\over2}
    -{\rm e}^{-i\gamma_{--0}(t)}\mathrm{sin}^2{\theta_0\over2}
  \right)
  \mathrm{sin}{\theta_1}
  \mathrm{cos}^2{\theta_2\over2}
\nonumber\\
&+&
  \sum_j
  {\rm e}^{i[ (\mathbf{q}-\mathbf{k}_2)\cdot\mathbf{r}_j 
             +(\nu_j-\omega)(t-t_2)]}
  {\rm e}^{-\Gamma_{-00}(t)}
  \nonumber\\
  &{}&\times
  {1\over2}
  \left[
     \left( {\rm e}^{ i\gamma_{-00}^+(t)}
            \mathrm{cos}^2{\theta_0\over2}
           -{\rm e}^{-i\gamma_{-00}^+(t)}
            \mathrm{sin}^2{\theta_0\over2}
     \right)
     \mathrm{cos}^2{\theta_1\over2}
     +
     \left( {\rm e}^{ i\gamma_{-00}^-(t)}
            \mathrm{sin}^2{\theta_0\over2}
           -{\rm e}^{-i\gamma_{-00}^-(t)}
            \mathrm{cos}^2{\theta_0\over2}
     \right)
     \mathrm{sin}^2{\theta_1\over2}
  \right]
  \mathrm{sin}{\theta_2}
\nonumber\\
&-&
  \sum_j
  {\rm e}^{i\{ [\mathbf{q} - (2\mathbf{k}_1-\mathbf{k}_0) ]
              \cdot\mathbf{r}_j 
             +(\nu_j-\omega)(t-t_2 + \Delta_1-\Delta_0)\} }
  {\rm e}^{-\Gamma_{--+}(t)}
  {1\over2}
  \mathrm{sin}{\theta_0}
  \mathrm{sin}^2{\theta_1\over2}
  \mathrm{cos}^2{\theta_2\over2}
\nonumber\\
&-&
  \sum_j
  {\rm e}^{i\{ [\mathbf{q} - (2\mathbf{k}_2-\mathbf{k}_0) ]
              \cdot\mathbf{r}_j 
             +(\nu_j-\omega)(t-t_2 - \Delta_1-\Delta_0)\} }
  {\rm e}^{-\Gamma_{-++}(t)}
  {1\over2}
  \mathrm{sin}{\theta_0}
  \mathrm{cos}^2{\theta_1\over2}
  \mathrm{cos}^2{\theta_2\over2}
\nonumber\\
&+&
  \sum_j
  {\rm e}^{i\{ [\mathbf{q} - (2\mathbf{k}_2-\mathbf{k}_1) ]
              \cdot\mathbf{r}_j 
             +(\nu_j-\omega)(t-t_2 - \Delta_1) \} }
  {\rm e}^{-\Gamma_{-+0}(t)}
  {1\over2}
  \left(
    -{\rm e}^{ i\gamma_{-+0}(t)}\mathrm{cos}^2{\theta_0\over2}
    +{\rm e}^{-i\gamma_{-+0}(t)}\mathrm{sin}^2{\theta_0\over2}
  \right)
  \mathrm{sin}{\theta_1}
  \mathrm{sin}^2{\theta_2\over2}
\nonumber\\
&-&
  \sum_j
  {\rm e}^{i\{ [\mathbf{q} 
               - (\mathbf{k}_2-\mathbf{k}_1+\mathbf{k}_0) ]
              \cdot\mathbf{r}_j 
             +(\nu_j-\omega)(t-t_2 + \Delta_0) \} }
      \left( 
           {\rm e}^{-\Gamma_{-0-}(t)}
           \mathrm{cos}\gamma_{-0-}(t) 
      \right)
  {1\over4}
  \mathrm{sin}{\theta_0}
  \mathrm{sin}{\theta_1}
  \mathrm{sin}{\theta_2}
\nonumber\\
&-&
  \sum_j
  {\rm e}^{i\{ [\mathbf{q} 
               - (\mathbf{k}_2+\mathbf{k}_1-\mathbf{k}_0) ]
              \cdot\mathbf{r}_j 
             +(\nu_j-\omega)(t-t_2 - \Delta_0) \} }
      \left( 
           {\rm e}^{-\Gamma_{-0+}(t)}
           \mathrm{cos}\gamma_{-0+}(t) 
      \right)
  {1\over4}
  \mathrm{sin}{\theta_0}
  \mathrm{sin}{\theta_1}
  \mathrm{sin}{\theta_2}
\nonumber\\
&+&
  \sum_j
  {\rm e}^{i\{ [\mathbf{q} 
               - (2\mathbf{k}_2-2\mathbf{k}_1+\mathbf{k}_0) ]
              \cdot\mathbf{r}_j 
             +(\nu_j-\omega)(t-t_2 - \Delta_1 + \Delta_0) \} }
      {\rm e}^{-\Gamma_{-+-}(t)} 
  {1\over2}
  \mathrm{sin}{\theta_0}
  \mathrm{sin}^2{\theta_1\over2}
  \mathrm{sin}^2{\theta_2\over2}. 
\end{eqnarray}
In the above equation, 
we have introduced the following quantities: 
\beq
\Gamma_{c_2 c_1 c_0}(t)
=
2\int_{0}^\infty d\Omega
I(\Omega)
\frac{f_{c_2 c_1 c_0}(\Omega,t)}{\Omega^2}
\mathrm{coth}
\left(  \frac{\hbar\Omega_l}{2k_\mathrm{B}T}  \right),
\eeq
\beq
f_{c_2 c_1 c_0}
=
\left\vert  
c_2 a^{(2)} + c_1 a^{(1)} + c_0 a^{(0)} 
\right\vert^2,
\quad (c_j\in\{+,-,0\})
\eeq
\beq
a^{(2)}(\Omega)
=
{\rm e}^{i\Omega t_2}-{\rm e}^{i\Omega t}, 
\eeq
\beq
a^{(1)}(\Omega)
=
{\rm e}^{i\Omega t_1}-{\rm e}^{i\Omega t_2}, 
\eeq
\beq
a^{(0)}(\Omega)
=
{\rm e}^{i\Omega t_0}-{\rm e}^{i\Omega t_1},   
\eeq
\beq
\gamma_{--0}(t)
=
4\int_{0}^\infty d\Omega
\frac{I(\Omega)}{\Omega^2}
\mathrm{Im}
\left\{  
   [a^{(2)}(\Omega)+a^{(1)}(\Omega)] a^{(0)}(\Omega)^\ast 
\right\},
\eeq
\beq
\gamma_{-+0}(t)
=
4\int_{0}^\infty d\Omega
\frac{I(\Omega)}{\Omega^2}
\mathrm{Im}
\left\{  
   [a^{(2)}(\Omega)-a^{(1)}(\Omega)] a^{(0)}(\Omega)^\ast 
\right\},
\eeq
\beq
\gamma_{-00}^+(t)
=
4\int_{0}^\infty d\Omega
\frac{I(\Omega)}{\Omega^2}
\mathrm{Im}
\left\{  
   a^{(2)}(\Omega)[a^{(1)}(\Omega)+a^{(0)}(\Omega)]^\ast 
\right\},
\eeq
\beq
\gamma_{-00}^-(t)
=
4\int_{0}^\infty d\Omega
\frac{I(\Omega)}{\Omega^2}
\mathrm{Im}
\left\{  
   a^{(2)}(\Omega)[a^{(1)}(\Omega)-a^{(0)}(\Omega)]^\ast 
\right\},
\eeq
and 
\beq
\gamma_{-0-}(t)
=
\gamma_{-0+}(t)
=
4\int_{0}^\infty d\Omega
\mathrm{Im}
\left[  
   a^{(2)}(\Omega)a^{(1)}(\Omega)^\ast 
\right]. 
\eeq
Here we are interested in the two signals 
observed in the directions of 
$2\mathbf{k}_2 - \mathbf{k}_0$ 
and 
$2\mathbf{k}_2 -2\mathbf{k}_1 + \mathbf{k}_0$.  
The former corresponds to the photon echo signal 
in the four wave mixing scheme 
which is not affected by the first pulse 
$(\mathbf{k}_1, \theta_1)$ at $t_1$.
On the other hand, 
the latter corresponds to the photon echo signal 
in the six wave mixing scheme, 
and contains the effect 
caused by the $(\mathbf{k}_1, \theta_1)$ pulses, 
respectively. 
The signal intensities are given by  
\beq\label{signal_4wmx}
I_{2\mathbf{k}_2 - \mathbf{k}_0}(t)
=
{1\over4}
{\rm sin}^2\theta_0
{\rm cos}^4\frac{\theta_1}{2}
{\rm sin}^4\frac{\theta_2}{2}
{\rm exp}[-2\Gamma_{-++}(t)
          -(t-t_2-\Delta_1-\Delta_0)^2 \delta_\mathrm{B}^2], 
\eeq
and 
\beq\label{signal_6wmx}
I_{2\mathbf{k}_2 -2\mathbf{k}_1 + \mathbf{k}_0}(t)
=
{1\over4}
{\rm sin}^2\theta_0
{\rm sin}^4\frac{\theta_1}{2}
{\rm sin}^4\frac{\theta_2}{2}
{\rm exp}[-2\Gamma_{-+-}(t)
          -(t-t_2-\Delta_1+\Delta_0)^2 \delta_\mathrm{B}^2], 
\eeq
where the inhomogeneous broadening distribution of 
Eq. (\ref{Inhomo_broadening_distribution}) has been assumed.

In the following we assume that 
$\theta_0=\theta_1=\theta_2=\pi/2$.  
This pulse area is recently achievable in the laboratory 
for common semiconductor quantum well and dot systems, 
such as GaSe and GaAs. 
Note that 
the signal intensities observed at 
$2\mathbf{k}_2 - \mathbf{k}_0$ 
and
$2\mathbf{k}_2 -2\mathbf{k}_1 + \mathbf{k}_0$
result from the qubits 
that effectively evolve under the 
($\mathbf{k}_0$, $\pi/2$ ; $\mathbf{k}_1$, $\pi$) 
and 
($\mathbf{k}_0$, $\pi/2$ ; $\mathbf{k}_1$, $\pi$ ; 
$\mathbf{k}_2$, $\pi$)
pulse sequence. 
The two signal intensities have the same weight factors. 
So we can directly compare the signal intensities observed 
at the two directions 
$I_{2\mathbf{k}_2 - \mathbf{k}_0}(t)$
and 
$I_{2\mathbf{k}_2 -2\mathbf{k}_1 + \mathbf{k}_0}(t)$ 
with each other. 
The difference between them is due to the effect 
by the first $\pi$ pulse, 
which is what should be confirmed experimentally. 
One suitable way to see this difference is 
to sweep $t_1$ by fixing $t_2$ such that 
the signal intensity $I_{2\mathbf{k}_2 - \mathbf{k}_0}(t)$ 
remains constant 
while 
$I_{2\mathbf{k}_2 -2\mathbf{k}_1 + \mathbf{k}_0}(t)$ varies.

\subsection{Homogeneous case}
\label{Weak:Homogeneous case}

\begin{figure} 
\begin{center} 
\includegraphics[width=17cm]{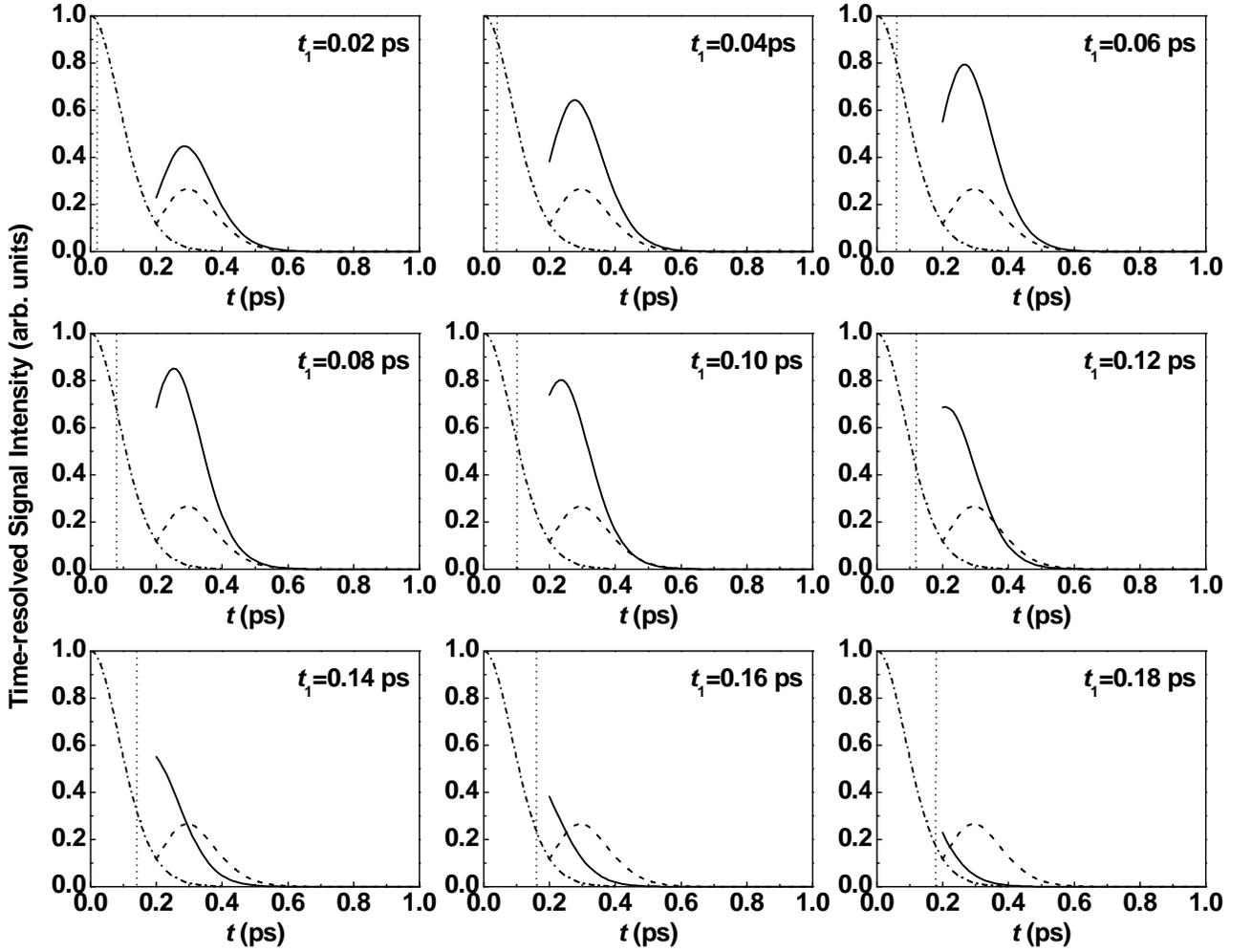} 
\end{center} 
\caption{
\label{wtmrslvohmic100kwc4a1} 
The intensities of the time-resolved six wave mixing signals 
in the case of 
homogeneous qubits and the ohmic reservoir model with 
$\alpha=0.1$, $\Omega_\mathrm{c}=4$\,meV,and $T=100$\,K. 
The thick solid line corresponds to the signal intensity 
$I_{2\mathbf{k}_2 -2\mathbf{k}_1 + \mathbf{k}_0}(t)$, 
while 
the thick dashed line to 
$I_{2\mathbf{k}_2 - \mathbf{k}_0}(t)$. 
The one dotted line represents the free induction decay signal. 
The time of the second pulse is added at $t_2=0.2$\,ps. 
The time of the first pulse 
$t_1=0.02, 0.04, ...,$ and 0.18\,ps is indicated 
by the vertical dashed line. 
}
\end{figure}

Let us start with the case of homogeneous qubits 
($\delta_\mathrm{B}=0$). 
Fig. \ref{wtmrslvohmic100kwc4a1} 
shows the signals 
$I_{2\mathbf{k}_2 -2\mathbf{k}_1 + \mathbf{k}_0}(t)$ 
(the solid line) 
and 
$I_{2\mathbf{k}_2 - \mathbf{k}_0}(t)$ 
(the dashed line) 
for various values of $t_1$ setting $t_2=0.2$\,ps. 
We have set $t_0=0$ hereafter. 
As a reference, the free induction decay signal 
$I_{\mathbf{k}_0}(t)$ 
is also shown by the one dotted line. 
The reservoir model is assumed to be the ohmic model with 
$\alpha=0.1$, $\Omega_\mathrm{c}=4$\,meV,and $T=100$\,K. 
The second pulse is added at $t_2=0.2$\,ps. 
The time of the first pulse 
$t_1=0.02, 0.04, ..., \mathrm{and}\ 0.18$\,ps is indicated 
by the vertical dashed line. 
The effect of decoherence suppression 
by the $(\mathbf{k}_1, \theta_1)$ pulse is seen 
for $t_1=0.02, 0.04, ..., \mathrm{and}\ 0.12$\,ps 
as the difference between the solid and dashed lines. 

\begin{figure} 
\begin{center} 
\includegraphics[width=17cm]{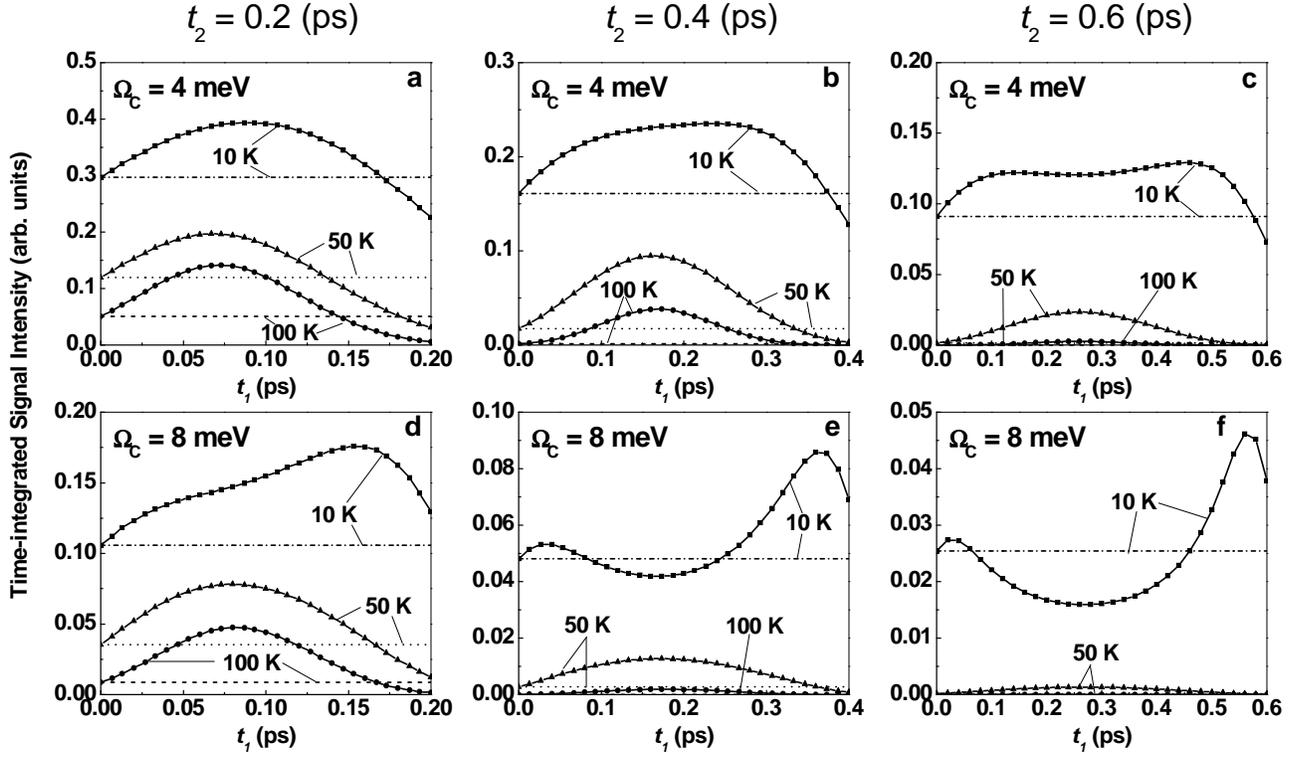} 
\end{center} 
\caption{
\label{wintohmicwc4wc8a1} 
The time-integrated signal intensities 
as a function of the first pulse $t_1$ 
with $t_2$ fixed. 
The homogeneous qubits and the ohmic reservoir model 
are assumed. 
The details are explained in the text. 
}
\end{figure}

When 
the signal $I_{2\mathbf{k}_2 -2\mathbf{k}_1 + \mathbf{k}_0}(t)$ 
is greater than 
$I_{2\mathbf{k}_2 - \mathbf{k}_0}(t)$ 
for almost all regions of $t>t_2$, 
(the cases of 
$t_1=0.1, 0.2, ..., \mathrm{and}\ 0.6$\,ps 
in Fig. \ref{wtmrslvohmic100kwc4a1}), 
one may use the time-integrated signal intensity, 
i.e., 
the total energy of the emitted radiation 
\beq
I_\mathbf{q}^\mathrm{int} 
=
\int_{t_2}^\infty dt I_\mathbf{q}(t) 
\eeq
to measure the effect of decoherence suppression. 
In fact, 
the measurement of the time-integrated signal intensity 
is much easier than that of 
the time resolved signal intensity $I_\mathbf{q}(t)$. 
Fig. \ref{wintohmicwc4wc8a1} shows 
the time-integrated signal intensity as a function of $t_1$, 
$I_\mathbf{q}^\mathrm{int}(t_1)$, 
fixing the second pulse time $t_2$. 
The upper three viewgraphs correspond to 
the case of the ohmic reservoir with 
$\alpha=0.1$ and $\Omega_\mathrm{c}=4$\,meV. 
$t_2$=0.2, 0.4, 0.6\,ps, from the left. 
Each viewgraph contains the cases of three kinds of temperature 
$T=$10, 50, and 100\,K. 
The lower three viewgraphs are the same as the upper 
but the higher cutoff frequency 
$\Omega_\mathrm{c}=8$\,meV. 
Since the signal at the direction 
$2\mathbf{k}_2 - \mathbf{k}_0$ 
is not affected by the first pulse, 
the time-integrated signal intensity 
as a function of $t_1$, 
$I_{2\mathbf{k}_2 - \mathbf{k}_0}^\mathrm{int}(t_1)$, 
results in a flat horizontal line. 
On the other hand, 
$I_{2\mathbf{k}_2 -2\mathbf{k}_1 + \mathbf{k}_0}
^\mathrm{int}(t_1)$
shows peak structure at certain time(s) of $t_1$. 
For the lower cutoff frequency $\Omega_\mathrm{c}$ and 
higher temperature, 
$I_{2\mathbf{k}_2 -2\mathbf{k}_1 + \mathbf{k}_0}
^\mathrm{int}(t_1)$ 
becomes larger than 
$I_{2\mathbf{k}_2 - \mathbf{k}_0}^\mathrm{int}(t_1)$ 
and is peaked around $t=t_2/2$ or an earlier time. 

\begin{figure} 
\begin{center} 
\includegraphics[width=6cm]{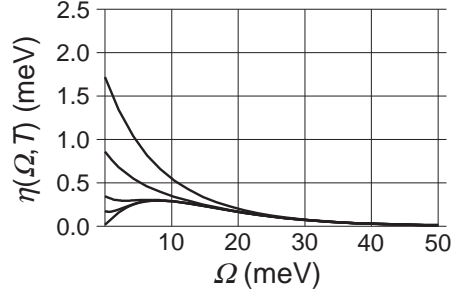} 
\end{center} 
\caption{
\label{thermalizedboson_ohmic} 
The thermalized boson factor 
$\eta(\Omega,T)$ 
in the case of the ohmic reservoir model with 
$\Omega_{\rm c}$=8\,meV and $\alpha=0.1$. 
The five curves correspond to the temperatures, 
from the bottom, 1K, 10K, 20K, 50K, and 100K.  
}
\end{figure}

For the higher cutoff frequency $\Omega_\mathrm{c}$ and 
lower temperature 
(the case of 10\,K 
in Figs. \ref{wintohmicwc4wc8a1}e and f), 
on the other hand, 
there appears a time region of $t_1$ 
where 
$I_{2\mathbf{k}_2 -2\mathbf{k}_1 + \mathbf{k}_0}
^\mathrm{int}(t_1)$ 
becomes smaller than 
$I_{2\mathbf{k}_2 - \mathbf{k}_0}^\mathrm{int}(t_1)$, 
i.e. 
the first pulse $(\mathbf{k}_1, \theta_1)$ fails 
in suppressing the decoherence 
and 
rather accelerates it. 
The origin of this phenomenon is the {\it in-phase} coupling 
between a qubit and the reservoir bosons 
caused by the first (intermediate) pulse 
$(\mathbf{k}_1, \theta_1)$, 
as explained in section 
\ref{Pi:Homogeneous case}. 
In order to suppress the decoherence effectively, 
the pulse interval should be 
$\Delta<\pi/(2 \Omega_\mathrm{th})$, 
where 
$\Omega_\mathrm{th}$ is the characteristic frequency 
at which the thermalized boson factor defined by 
\beq\label{thermalized boson factor}
\eta(\Omega,T)
=
I(\Omega)
{\rm coth}(\frac{\hbar\Omega}{2k_{\rm B}T}),  
\eeq
is peaked. 
Note that the decoherence exponent 
in Eq. (\ref{DecoherenceExponent}) 
can be expressed as  
\beq
\Gamma(t)
=
\int_{0}^\infty d\Omega
\xi(\Omega,t)\eta(\Omega,T),  
\eeq
using $\eta(\Omega,T)$ and 
the time dependent factor defined by 
\beq
\xi(\Omega,t)
=
2\frac{f(\Omega,t)}{\Omega^2}. 
\eeq
The condition 
$\Delta<\pi/(2 \Omega_\mathrm{th})$ 
is not satisfied in the case of 10\,K 
in Figs. \ref{wintohmicwc4wc8a1}e and f. 
In fact, 
for the higher cutoff ferequency $\Omega_\mathrm{c}$ and 
lower temperature, 
$\Omega_\mathrm{th}$ becomes higher 
as seen in Fig. \ref{thermalizedboson_ohmic}. 

It is commonly encountered in semiconductor exciton systems 
that the decoherence is accelerated 
when an additional pulse is irradiated. 
This is usually due to that 
an additional pulse increases the number of excitons, 
which in turn enhances the exciton-exciton interaction,  
and increases the number of decoherence channels. 
In the present model, however, 
the interaction between the qubits themselves is not 
taken into account. 
Therefore the decoherence acceleration effect purely 
comes from the coherent interaction between a qubit 
and the reservoir bosons. 
More remarkably 
in the case of 10\,K 
in Figs. \ref{wintohmicwc4wc8a1}e and f, 
$I_{2\mathbf{k}_2 -2\mathbf{k}_1 + \mathbf{k}_0}
^\mathrm{int}(t_1)$ 
exceeds 
$I_{2\mathbf{k}_2 - \mathbf{k}_0}^\mathrm{int}(t_1)$ 
again, and forms a peak for larger $t_1$. 
In this region, 
the {\it out-of-phase} coupling 
between a qubit and the reservoir bosons dominates, 
and the first pulse $(\mathbf{k}_1, \theta_1)$ succeeds 
in suppressing the decoherence. 
This kind of cross over from the decohernce acceleration 
to the suppression is a peculiar aspect of the present model, 
which is expected at low temperatures, 
and must be interesting in its own light 
to be investigated experimentally.

\subsection{Inhomogeneous case}
\label{Weak:Inhomogeneous case}

\begin{figure} 
\begin{center} 
\includegraphics[width=12cm]{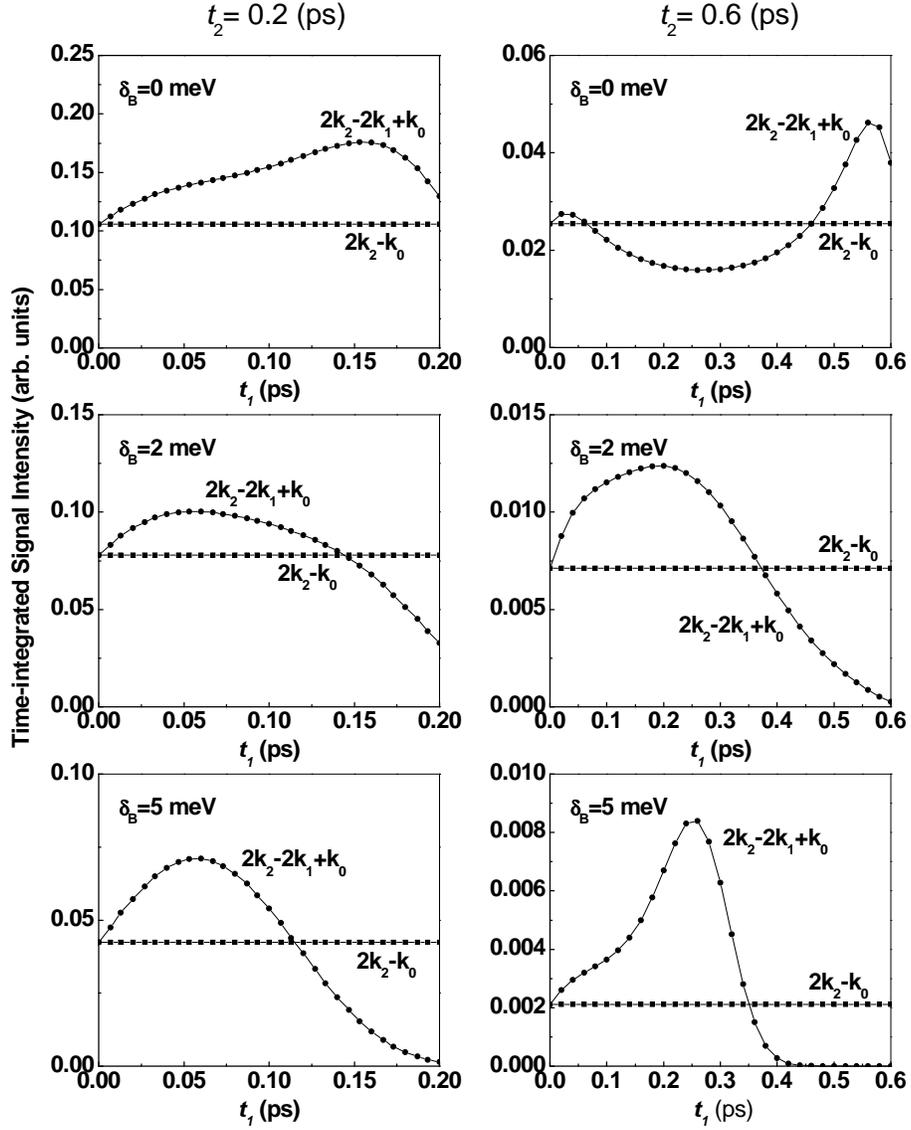} 
\end{center} 
\caption{
\label{wintohmicinhm10kwc8a1} 
The time-integrated signal intensities 
as a function of the first pulse $t_1$ 
with $t_2$ fixed. 
The ohmic reservoir model is assumed 
with the parameters $\alpha=0.1$ and 
$\Omega_\mathrm{c}=$8\,meV. 
The top figures correspond to the homogeneous case, 
while 
the middle and bottom ones to the inhomogeneous case 
with 
$\delta_\mathrm{B}$=2\,meV 
and 
$\delta_\mathrm{B}$=5\,meV, 
respectively. 
}
\end{figure}
\begin{figure} 
\begin{center} 
\includegraphics[width=6cm]{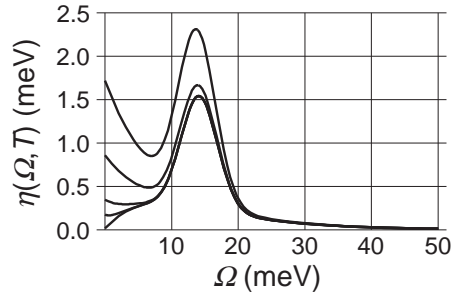} 
\end{center} 
\caption{
\label{thermalizedboson_gaussian} 
The thermalized boson factor 
$\eta(\Omega,T)$ 
in the case of the Gaussian-ohmic reservoir model with 
$\Omega_\mathrm{p}=$13\,meV, 
$\gamma_\mathrm{p}=$4\,meV, 
$\alpha_\mathrm{p}=0.05$, 
$\Omega_\mathrm{c}=$8\,meV, 
and 
$\alpha=0.1$. 
The five curves correspond to the temperatures, 
from the bottom, 1K, 10K, 20K, 50K, and 100K.  
}
\end{figure}
\begin{figure} 
\begin{center} 
\includegraphics[width=12cm]{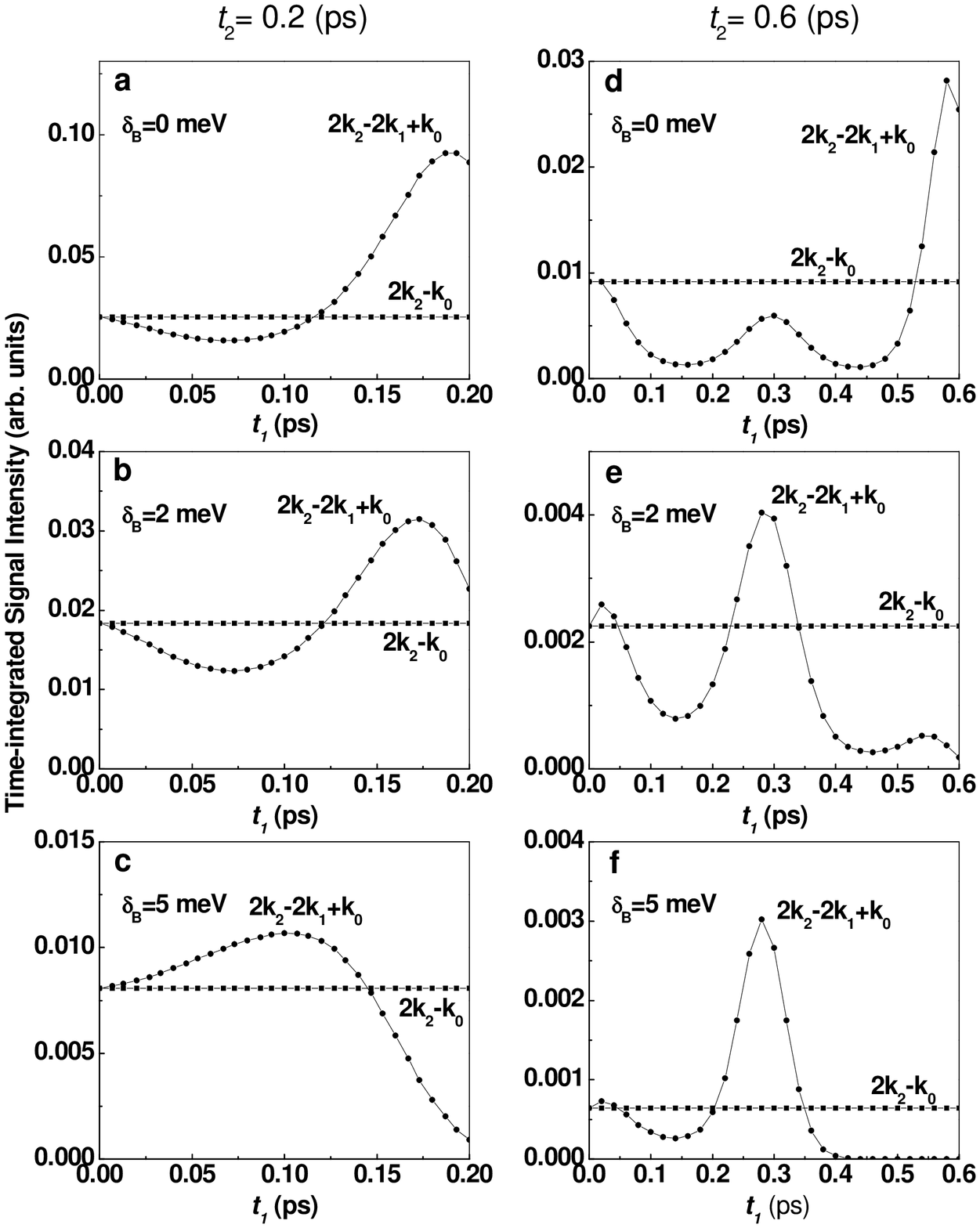} 
\end{center} 
\caption{
\label{wintgaussinhm10kwc8a1wp13gp4ap05} 
The time-integrated signal intensities 
as a function of the first pulse $t_1$ 
with $t_2$ fixed 
in the case of the Gaussian-ohmic reservoir model 
with the parameters 
$\Omega_\mathrm{p}=$13\,meV, 
$\gamma_\mathrm{p}=$4\,meV, 
$\alpha_\mathrm{p}=0.05$,
$\Omega_\mathrm{c}=$8\,meV, 
and 
$\alpha=0.1$. 
The top figures correspond to the homogeneous case, 
while 
the middle and bottom ones to the inhomogeneous case 
with 
$\delta_\mathrm{B}$=2\,meV 
and 
$\delta_\mathrm{B}$=5\,meV, 
respectively. 
}
\end{figure}
Now let us consider the inhomogeneous broadening effect 
in a bulk ensemble of qubits. 
Fig. \ref{wintohmicinhm10kwc8a1} shows 
how the time-integrated signal intensities change 
when the inhomogeneous broadening effect is introduced
into the ohmic reservoir model. 
From the upper viewgraphs 
the inhomogeneous broadening effect is taken as 
$\delta_\mathrm{B}$=0, 2, and 5\,meV. 
The reservoir parameters are 
$\alpha=0.1$ and $\Omega_\mathrm{c}=8$\,meV. 
The time of the second pulse $t_2$ is 
0.2\,ps and 0.6\,ps for the left and right viewgraphs, 
respectively. 
It should be noted that 
the inhomogeneously broadened signals 
$I_{2\mathbf{k}_2 -2\mathbf{k}_1 + \mathbf{k}_0}(t)$ 
and 
$I_{2\mathbf{k}_2 - \mathbf{k}_0}(t)$
appear as the echos around the times 
$t=t_2+\Delta_1-\Delta_0=2(t_2-t_1)$ 
and 
$t=t_2+\Delta_1+\Delta_0=2 t_2$, 
respectively. 
While the temporal position of the echo signal 
$I_{2\mathbf{k}_2 - \mathbf{k}_0}(t)$ does not depend on $t_1$, 
the echo signal 
$I_{2\mathbf{k}_2 -2\mathbf{k}_1 + \mathbf{k}_0}(t)$ 
appears earlier as $t_1$ gets longer delay. 
In particular, 
when the time $t_1$ exceeds $t_2/2$ 
such that $\Delta_1<\Delta_0$, 
the echo signal reduces exponentially. 
As seen, 
it is still possible to observe the decoherence suppression  
effect, 
i.e. that 
$I_{2\mathbf{k}_2 -2\mathbf{k}_1 + \mathbf{k}_0}
^\mathrm{int}(t_1)$ 
becomes larger than 
$I_{2\mathbf{k}_2 - \mathbf{k}_0}^\mathrm{int}(t_1)$ 
even in the presence of the inhomogeneous broadening effect. 
On the other hand, 
the decoherence acceleration effect seen in the right top 
viewgraph is rapidly smeared out 
as the amount of inhomogeneous broadening $\delta_\mathrm{B}$ 
increases. 
%
%
%
%

The decoherence acceleration effect can be observed
more clearly in the Gaussian-ohmic reservoir model, 
which has a sharper spectral density around 
a characteristic frequency 
$\Omega_\mathrm{p}$ 
than that in the ohmic reservoir model. 
Fig. \ref{thermalizedboson_gaussian} shows 
the thermalized boson factor $\eta(\Omega, T)$. 
Fig. \ref{wintgaussinhm10kwc8a1wp13gp4ap05} 
shows the time integrated signal intensities 
for the Gaussian-ohmic reservoir model with the parameters 
$\Omega_\mathrm{p}=$13\,meV, 
$\gamma_\mathrm{p}=$4\,meV, 
$\alpha_\mathrm{p}=0.05$, 
$\Omega_\mathrm{c}=$8\,meV, 
and 
$\alpha=0.1$. 
From the upper viewgraphs 
the inhomogeneous broadening effect is taken as 
$\delta_\mathrm{B}=$0, 2, and 5meV.
The time $t_2$ is 0.2\,ps and 0.6\,ps 
for the left and right viewgraphs,
respectively.
Among the six viewgraphs, 
we would like to pay attension to 
Figs. \ref{wintgaussinhm10kwc8a1wp13gp4ap05}b, e, and f. 
In this case, 
$I_{2\mathbf{k}_2 -2\mathbf{k}_1 + \mathbf{k}_0} 
^\mathrm{int}(t_1)$ 
exhibits the decoherence acceleration effect first, 
the decoherence suppression effect then, 
as a function of $t_1$. 
Such a cross over is a very characteristic evidence 
of the coherent qubit-reservoir interaction 
in the non-Markovian region. 
More importantly, this behavior can be seen even under the 
realistic amount of the inhomogeneous broadening effect 
such as $\delta_\mathrm{B}=$5meV.
This may appeal to experimental investigation 
on the coherent qubit-reservoir interaction.

%
%
%
%
%

\section{Concluding remark}
\label{Concluding remark}

We have developed a theory to analyze the decoherence 
in the qubit array system with 
the photon echo signals in the multiwave mixing configuration. 
We have presented how the decoherence suppression effect 
by the BB control with the $\pi$ pulses 
can be demonstrated in laboratory 
by using a bulk ensemble of excitons and 
optical pulses whose pulse area is even smaller than $\pi$. 
The key is 
to analyze the time-integated multiwave mixing signals 
diffracted into certain phase matching directions
from a bulk ensemble. 
We have given numerical examples in the six wave mixing 
configuration (three pulses), 
where 
the signals in two directions 
$2\mathbf{k}_2 - \mathbf{k}_0$ 
and 
$2\mathbf{k}_2 -2\mathbf{k}_1 + \mathbf{k}_0$ 
are compared with each other by sweeping the pulse intervals. 
Depending on the pulse interval conditions, 
both the decoherence suppression and acceleration effects 
take place as pointed out in earlier works. 
The cross over from one to the other is a clear evidence 
of the coherent qubit-reservoir interaction. 
We have shown that this cross over may be observed 
even under realistic inhomogeneous broadening. 
This encourages experimental investigations 
in qubit systems in solid state.

To understand this cross over, 
we have introduced the notions of 
{\it in-phase} and {\it out-of-phase} couplings 
between 
a qubit and the reservoir bosons. 
The decoherence process in the present model 
is mathematically described by 
the state evolution caused by 
the displacement operations on the reservoir boson modes, 
depending on the qubit states of $\hat\sigma_z$ components. 
The {\it in-phase} coupling means that 
the successive displacement operations between the optical 
pulses act additively, 
enhancing the entanglement 
between a qubit and the reservoir bosons. 
The {\it out-of-phase} coupling means that 
the successive displacement operations act as canceling out.

Which type of coupling dominates 
is determined by the relation between the pulse interval 
and the reservoir boson spectrum.  
Therefore by analyzing such behaviors systematically 
one may identify the reservoir characteristics. 
In particular, when analysis is made in the weak pulse 
configuration where we can observe many diffraction signals 
from various multiwave mixing channels simultaneously, 
we may increase the precision of parameter estimation 
by fitting as many diffracted signals as possible 
with a single theoretical model. 
Thus our theory will be useful to investigate 
the reservoir characteristics,  
which is an important first step to proceed 
quantum information processing with semiconductor excitons.

\acknowledgements

The authors would like to thank T. Kishimoto, 
C. Uchiyama and M. Ban 
for helpful discussions.

\end{document}